\documentclass[sigconf]{acmart}
\AtBeginDocument{%
  \providecommand\BibTeX{{%
    \normalfont B\kern-0.5em{\scshape i\kern-0.25em b}\kern-0.8em\TeX}}}

\usepackage{soul}
\usepackage{graphicx}
\usepackage{booktabs}
\usepackage{float}
\usepackage{overpic}
\usepackage{float}  
\usepackage{subfloat}
\usepackage{stfloats} 
\usepackage[skip=0.5\baselineskip]{caption}
\usepackage{bm}
\usepackage{amsmath}
\usepackage{booktabs}
\usepackage{multirow}
\usepackage{multicol}
\usepackage{enumitem}
\usepackage{subcaption}
\usepackage{threeparttable}
\usepackage{xspace}
\usepackage{color}
\usepackage[normalem]{ulem}
\usepackage{algorithmicx,algorithm}
\usepackage{algpseudocode}
\usepackage{algorithm,algpseudocode}
\usepackage{marvosym}   
\usepackage{colortbl}   
\usepackage{tcolorbox}  
\tcbuselibrary{breakable}   
\usepackage{wasysym}    
\usepackage[figuresright]{rotating} 
\newcommand{\name}{DRP\xspace}
\makeatletter
\def\@ACM@checkaffil{
    \if@ACM@instpresent\else
    \ClassWarningNoLine{\@classname}{No institution present for an affiliation}%
    \fi
    \if@ACM@citypresent\else
    \ClassWarningNoLine{\@classname}{No city present for an affiliation}%
    \fi
    \if@ACM@countrypresent\else
        \ClassWarningNoLine{\@classname}{No country present for an affiliation}%
    \fi
}
\makeatother

\usepackage{xspace}

\newcommand{\ie}{\emph{i.e.,}\xspace}

\newcommand{\eat}[1]{}

\copyrightyear{2025}
\acmYear{2025}
\setcopyright{acmlicensed}\acmConference[WWW '25]{Proceedings of the ACM
Web Conference 2025}{April 28-May 2, 2025}{Sydney, NSW, Australia}
\acmBooktitle{Proceedings of the ACM Web Conference 2025 (WWW '25), April
28-May 2, 2025, Sydney, NSW, Australia}
\acmDOI{10.1145/3696410.3714949}
\acmISBN{979-8-4007-1274-6/25/04}

\begin{document}
\begin{sloppypar}   



\title{Behavior Modeling Space Reconstruction for E-Commerce Search}
\author{Yejing Wang}
\email{yejing.wang@my.cityu.edu.hk}
\affiliation{
    \institution{City University of Hong Kong}
    \city{Hong Kong SAR}
    \country{China}
}

\author{Chi Zhang}
\affiliation{
    \institution{Harbin Engineering University}
    \city{Harbin}
    \country{China}
}
\author{Xiangyu Zhao}
\authornote{Corresponding author.}
\email{xianzhao@cityu.edu.hk}
\affiliation{
    \institution{City University of Hong Kong}
    \city{Hong Kong SAR}
    \country{China}
}

\author{Qidong Liu}
\affiliation{%
  \institution{Xi'an Jiaotong University \& \\ City University of Hong Kong}
  \city{Xi'an}
  \country{China}
}

\author{Maolin Wang}
\affiliation{
    \institution{City University of Hong Kong}
    \city{Hong Kong SAR}
    \country{China}
}

\author{Xuetao Wei}
\affiliation{
\institution{Southern University of Science and Technology}
\city{Shenzhen}
\country{China}
}

\author{Zitao Liu}
\affiliation{
\institution{Jinan University}
\city{Guang Zhou}
\country{China}
}

\author{Xing Shi}
\affiliation{
    \institution{Alibaba Cloud}
    \city{Hangzhou}
    \country{China}
}

\author{Xudong Yang}
\affiliation{
    \institution{Alibaba Cloud}
    \city{Hangzhou}
    \country{China}
}
\author{Ling Zhong}
\affiliation{
    \institution{Alibaba Cloud}
    \city{Hangzhou}
    \country{China}
}

\author{Wei Lin}
\affiliation{
    \institution{Alibaba Cloud}
    \city{Hangzhou}
    \country{China}
}

\renewcommand{\shortauthors}{Yejing Wang, et al.}

\begin{abstract}
Delivering superior search services is crucial for enhancing customer experience and driving revenue growth in e-commerce. Conventionally, search systems model user behaviors by combining user preference and query-item relevance statically, often through a fixed logical `and' relationship.
This paper reexamines existing approaches through a unified lens using both causal graphs and Venn diagrams, uncovering two prevalent yet significant issues: entangled preference and relevance effects, and a collapsed modeling space.
To surmount these challenges, our research introduces a novel framework, \name, which enhances search accuracy through two components to reconstruct the behavior modeling space. Specifically, we implement preference editing to proactively remove the relevance effect from preference predictions, yielding untainted user preferences. Additionally, we employ adaptive fusion, which dynamically adjusts fusion criteria to align with the varying patterns of relevance and preference, facilitating more nuanced and tailored behavior predictions within the reconstructed modeling space.
Empirical validation on two public datasets and a proprietary e-commerce search dataset underscores the superiority of our proposed methodology, demonstrating marked improvements in performance over existing approaches. The code is available at \url{https://github.com/Applied-Machine-Learning-Lab/DRP}.
\hspace*{\fill}

\end{abstract}


\maketitle

\section{Introduction}
E-commerce has revolutionized the retail industry by harnessing the capabilities of the World Wide Web to offer consumers round-the-clock access to a global array of products and services~\cite{jain2021overview,wang2023single,wang2023doctor,li2023imf,wang2022autofield,wang2023plate,wang2024gprec}. 
In e-commerce applications, the search functionality serves as a critical determinant, matching and ranking relevant items in response to user queries, thereby driving user engagement (e.g., clicks)~\cite{xu2024large,zhao2020whole,zhao2018recommendations,zhao2018deep}. 
Consequently, a fundamental objective of search model development is to accurately and comprehensively understand user behaviors (i.e., why users interact with certain presented items instead of others)~\cite{lin2022adafs,li2022gromov,zheng2022cbr,li2023hamur}. However, in practice, available data often only indicates whether a user interacted with an item, without providing explicit insights into the underlying motivations~\cite{liu2023multi,liu2023multimodal,liang2023mmmlp,lin2023autodenoise}. Thus, effectively modeling user behaviors remains one of the most significant technical challenges in e-commerce search optimization.


User behaviors in e-commerce search can be majorly attributed to the query-item relevance and the user preference on items~\cite{carmel2020irrlevant-amazon} since the behavior data usually consists of candidate items, users, and their queries for ongoing search sessions.  
Accordingly, previous studies can be divided into three research lines: relevance-only modeling, preference-only modeling, and the joint of both. A fundamental assumption of the first two approaches is that utilizing relevance or preferences provides sufficient information to model user behaviors and explains why users interact with particular items.
Relevance-only methods regard e-commerce search as a basic 
user-independent search problem and focus on capturing the relevance between queries and items. 
As such, established ad-hoc search methodologies, such as BM25~\cite{robertson2009probabilistic} and DSSM~\cite{huang2013learning}, can be effectively applied as the relevance-only strategy. Some researchers have suggested that relevance effects on user behaviors are highly personalized and advanced relevance-only methods by integrating user features for personalized search systems~\cite{ai2019explainableDREM,dai2023contrastive,bi2020transformerTEM}. While these methods adopt a search perspective that prioritizes relevance, an alternative viewpoint considers the query as an integral part of the user features, reflecting user intent for a particular search session. Under this assumption, recommendation models~\cite{guo2017deepfm,cheng2016wd,li2023strec,liu2024large,liu2024moe} that better capture user preferences for items are usually deployed as preference-only methods. 
The first two lines tend to model relevance or preference effects individually and are trained with behavior signals, suffering from the misaligned training~\cite{nguyen2020learning,liu2022rating,liu2023exploration}. We will further illustrate this mismatch as the concept of collapsed modeling space in Section~\ref{sec:pre_problems} based on a Venn diagram. 

Additionally, these models typically produce a single relevance or preference prediction, while neglecting either effect could lead to suboptimal outcomes, as highlighted by \cite{carmel2020irrlevant-amazon}. For example, prioritizing preference alone can create an echo chamber~\cite{ge2020understanding,zhang2022hierarchical,zhao2023embedding}, where favored yet irrelevant items are ranked higher than more relevant ones.
To address this, an instant solution is to include both effects for joint modeling. Most existing joint frameworks focus on developing pre-trained relevance models~\cite{xiao2019weakly, zan2023spm, chen2023beyond,yao2021learningpositionbias,jiang2020bert2dnn}. In these frameworks, the relevance model is initially trained on human-labeled datasets that consist of query-item relevance signals. Subsequently, they are integrated with the preference model for joint optimization by the behavior data and finally deployed in search systems. 
However, this approach still faces two challenges:
\textbf{i) Biased and collapsed modeling space.} Since the relevance model is trained before the preference model, the combined results may disproportionately favor relevant items, resulting in a biased modeling space. In addition, the current scoring formula is static, lacking the flexibility to adapt to the diverse patterns of samples with different relevance-preference statuses, finally collapsing the model space. We will illustrate this point with a Venn diagram in Section~\ref{sec:pre_problems}. 
\textbf{ii) Laborious data collection.} Pre-trained relevance models heavily depend on exhaustively collected human-labeled relevance data, making them difficult to apply universally across various e-commerce platforms, particularly for smaller companies that cannot afford such labor-intensive data collection.
\textit{Thus, there is a pressing need for a joint modeling framework that captures user behaviors within correct modeling spaces, while also possessing adaptive scoring capabilities without relying on human-labeled data.}

Furthermore, existing methods overlook the inherent influence of relevance on user preference as visualized by the causal graph in Figure~\ref{fig:CauVenn} (a)\footnote{More details will be provided in Section~\ref{sec:pre_problems}.}. This unaddressed influence can lead to impure preference prediction, further collapsing the modeling space and finally impairing the joint behavior modeling. 
To tackle these, this paper introduces \textbf{\name}, which models user behavior through \textbf{\underline{D}}isentangling \textbf{\underline{R}}elevance and \textbf{\underline{P}}reference effects, alongside an adaptive fusion formula. Specifically, \name begins with eliminating the relevance effect from preference predictions by editing corresponding representations, thus reconstructing the fine-grained modeling space. \name subsequently learns nuanced user behaviors, such as item clicks, based on the dual-level adaptive integration within the reconstructed modeling space. Our major contributions are as follows:
\begin{itemize}[leftmargin=*]

\item We build a theoretical framework based on causal graphs and Venn diagrams to comprehensively review existing behavior modeling approaches for e-commerce search.
\item We explicitly divide the behavior modeling space into six areas, based on which we point out the collapsed modeling space caused by three problems of existing methods. 
\item We introduce \name, effectively addressing the collapsed modeling space with disentangled relevance and preference effects, as well as dual-level adaptive fusion. Notably, \name operates without the need for human-labeled relevance data, making it deployable with just behavior data.
\item We perform extensive experiments on two large public datasets and one private e-commerce search dataset, demonstrating the superior performance of \name.

\end{itemize}
\section{Preliminary}

\subsection{Joint Modeling Framework}\label{sec:basic_frame}
In this section, we outline the notation and fundamental structure of the joint modeling framework employed in this paper. 

\begin{figure*}[t]
    \raggedleft
    \includegraphics[width=\linewidth]{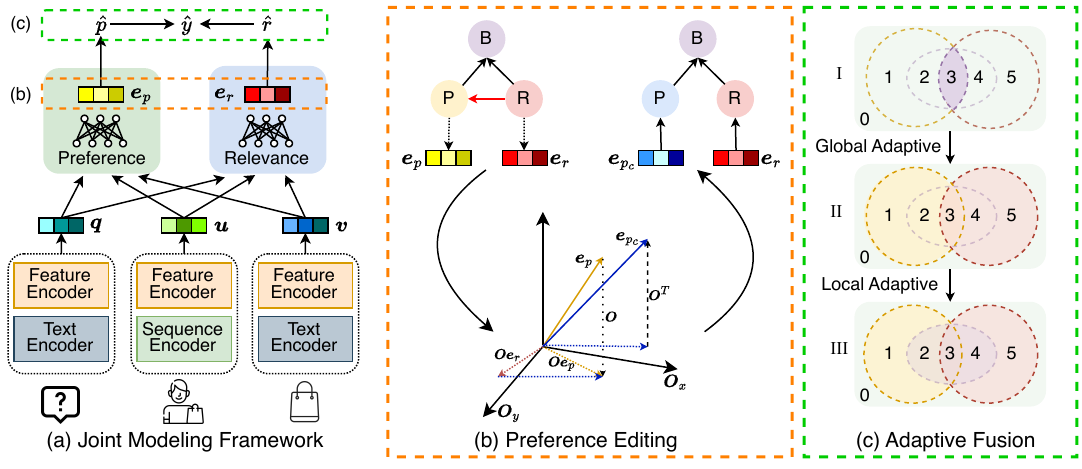}
    \caption{Framework Overview. The joint modeling framework is depicted in (a). The preference editing is represented in (b). Adaptive fusion is illustrated in (c).}
    \label{fig:Frame}
    \vspace{-4mm}
\end{figure*}

The structure is visually represented in Figure~\ref{fig:Frame}(a). Typically, the textual information from the input query is transformed into dense representations using a text encoder. These representations, along with additional query features such as query frequency, are then synthesized into the query embedding $\boldsymbol{q}$. Likewise, the text associated with items—such as titles or product names—is also converted with the same text encoder and then combined with other item features as the item embedding $\boldsymbol{v}$. For user data, we integrate users' static features as `Activity Degree' and encoded behavioral sequences features as `Clicked Items' into the user embedding $\boldsymbol{u}$. 

Then, the relevance prediction $\hat r$ is generated through the relevance model, denoted as $\mathrm{RM}$. For ad-hoc search methods~\cite{huang2013learning,robertson2009probabilistic}, it can be mathematically expressed as: 
\begin{gather}
    \hat{r} = \mathrm{RM}(\boldsymbol{q},\boldsymbol{v}) \label{eq:RM}
\end{gather}
In the case of personalized search methods~\cite{bi2021learningREM,ai2019zeroZAM}, the relevance model incorporates the user embedding to facilitate personalization, yielding $\hat{r} = \mathrm{RM}_{u}(\boldsymbol{q},\boldsymbol{v})$. 

The user preference for a specific item under the query can likewise be predicted using the preference model $\mathrm{PM}$:
\begin{gather}
    \hat{p} = \mathrm{PM}(\boldsymbol{q},\boldsymbol{v},\boldsymbol{u}) \label{eq:PM}
\end{gather}
The query embedding $\boldsymbol{q}$ is consistently provided as input to the preference model, as it encapsulates the user’s intent for the current session—an essential aspect of preference modeling.

Finally, the relevance and preference predictions are integrated to produce the final score, typically calculated via multiplication::
\begin{gather}
    \hat{y} = \hat{r}^\delta\cdot\hat{p} \label{eq:fixed_scoring}
\end{gather}
where $\hat{y}$ represents the final behavior prediction, $\delta$ is a hyper-parameter that adjusts the balance between relevance and preference. For relevance-only modeling, we assume $\hat{p}\equiv 1$, while for preference-only modeling, $\hat{r}\equiv 1$.

\begin{figure}[t]
    \raggedleft
    \includegraphics[width=\linewidth]{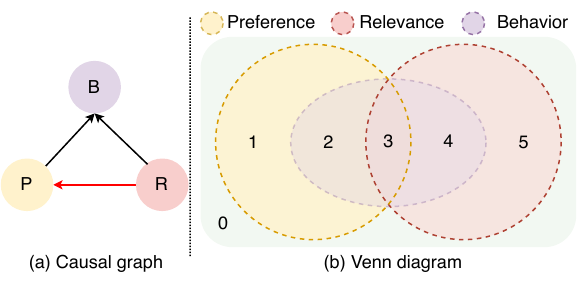}
    \caption{Causal graph and Venn diagram.}
   \vspace{-3mm}
    \label{fig:CauVenn}
\end{figure}

\begin{table}[t]
\centering
\caption{Area indicators for Venn diagram.}
\label{tab:Area_num}
\begin{tabular}{@{}ccccccc@{}}
\toprule
Area\# & 0 & 1 & 2 & 3 & 4 & 5 \\ \midrule
P      & 0 & 1 & 1 & 1 & 0 & 0 \\
R      & 0 & 0 & 0 & 1 & 1 & 1 \\
B      & 0 & 0 & 1 & 1 & 1 & 0 \\ \bottomrule 
\end{tabular}
\vspace{-4mm}
\end{table}

\subsection{Collapsed Modeling Space Problem}\label{sec:pre_problems}
This section will first introduce the causal graph and Venn diagram for behavior modeling in e-commerce search. Then we point out the collapsed modeling space of existing methods.

The causal graph is illustrated in Figure~\ref{fig:CauVenn}(a), where $B$ represents user behavior—such as clicks and purchases—while $P$ and $R$ denote preference and relevance, respectively. In this graph, the black arrows $P\to B$ and $R\to B$ indicate the preference and relevance effect on user behaviors. It is clear that users would respond to the returned items based on these effects: they may click on items with either high relevance or strong preference.
Besides, the preference is also influenced by relevance (the red arrow $R\to P$). 
When the item is highly relevant to the user query, the preference score may increase, reflecting a stronger affinity for the item in the context of the specific query—even if the user typically does not favor it. \textbf{For instance}, a user who prefers a formal dressing style may buy `Nike Shoes' because they are highly relevant when searching for `Sports Shoes'. In this case, the model might learn the user’s preference for `Nike Shoes' or the Nike brand although the user generally prefers formal dressing brands. 
This inflated preference effect can finally harm the search system. In the same example, the next time this user looks for a dress for everyday outfits, where sports styles aren't favored, the model might still suggest `Nike Dress' because it is relevant to the search and is consistent with the biased preference on `Nike' learned from the earlier session.


Existing joint modeling frameworks primarily concentrate on the effects of relevance and preference on user behaviors ($P\to B$ and $R\to B$), overlooking the influence of relevance on preference signals ($R\to P$). Based on the assumption that only $P\to B$ and $R\to B$ exist in Figure~\ref{fig:CauVenn}(a), we visualize the behavior modeling space with a Venn diagram in Figure~\ref{fig:CauVenn}(b). In this diagram, the green rectangle encompasses all behavior data, the yellow circle represents user preferences, the red circle denotes positive relevance effects, and the purple ellipse highlights positive behaviors. We label six distinct areas with numbers 0-5, with corresponding preference, relevance, and behavior signals detailed in Table~\ref{tab:Area_num}. For example, in click modeling, Area \#0 corresponds to the situation where $P=0,R=0,B=0$. This indicates that the item is neither relevant to the query nor favored by users, and with a negative click signal. We divide the Venn diagram into six areas as $P=0,R=0,B=1$ and $P=1,R=1,B=0$, contradicting our causal assumption that the behavior is only attributed to the relevance and preference.


We suggest that current works typically suffer from the collapsed modeling space.
For relevance-only or preference-only methods~\cite{ai2017learningHEM,wang2021masknet}, the single relevance and preference effects are trained with the behavior signals, leading to the modeling space mismatch. Specifically, the relevance model aims to learn the relevance effect (Area\#3-5), while its training signals are derived from the behavior space (Area\#2-4). Consequently, the relevance model may misclassify Area\#2 as relevant and Area\#5 as irrelevant, contrary to the actual situation. This similar misalignment also occurs in preference-only methods, where Area\#4 is incorrectly perceived as a preference and Area\#1 as the dislike. 

Besides, there are three problems for joint models based on the modeling space.
\textbf{Problem 1:} The neglect of the relevance influence on the preference of joint modeling methods can collapse the modeling space, as it may ruin the preference modeling~\cite{guo2020survey,liu2024largesurvey}. In the worst case, we can imagine that only relevant items can be preferred by users, where Area\#1\&2 vanish. 
\textbf{Problem 2:} Existing joint models~\cite{hong2024print,zan2023spm} usually integrate two effects with a fixed formula as in Equation~\eqref{eq:fixed_scoring}. However, this static integration falls short of capturing the complexities of modeling space, which can exhibit diverse relevance-preference patterns. Equation~\eqref{eq:fixed_scoring} only learns the positive behavior with positive preference and relevance effects. As a result, it would only distinguish Area\#0\&3, missing the pattern of Area\#2\&4 and Area\#1\&5. 
\textbf{Problem 3:} Current optimization scheme that trains the relevance model before integrating it with the preference model~\cite{ zan2023spm, chen2023beyond} may also collapse the modeling space. Under this pipeline, the joint model first identifies the relevant space (Area\#3-5). During the subsequent training of preference effect, the model may prioritize relevant items over preferred items in any situation. For instance, comparing Area\#4 and Area\#2 reveals that Area\#4 can be learned more easily due to the pre-trained relevance knowledge, while the model may miss Area\#2 because it contradicts this prior knowledge, e.g., irrelevant but clicked. Additionally, when we compare Area\#2 and Area\#3, it becomes evident that prior relevance knowledge can disrupt preference model training. Both Area\#2 and Area\#3 exhibit the same preference and behavioral signals ($P=1,B=1$), yet opposing relevance.

\section{Method}

\subsection{Overview}
This section briefly introduces the overall process of \name with 
the framework overview illustrated in Figure~\ref{fig:Frame}. 

Figure~\ref{fig:Frame} (a) presents the basic structure of the joint modeling frameworks introduced in Section~\ref{sec:basic_frame}. The joint model first obtains the query and item representations, $\boldsymbol{q}$ and $\boldsymbol{r}$, using the text encoder and feature encoder, along with the user representation $\boldsymbol{u}$ from the sequence encoder and feature encoder. Subsequently, relevance and preference effects are predicted based on these representations, as described in Equation~\eqref{eq:RM} and Equation~\eqref{eq:PM}. These predictions are then merged to generate the behavior prediction, as shown in Equation~\eqref{eq:fixed_scoring}. The proposed \name operates on the last layer representation of the preference model (within the orange dashed block) and the scoring formula for behavior prediction (within the green dashed block), without constraining the underlying feature encoders or backbone models.
Specifically, we design a preference editing framework to eliminate the relevance influence from the preference prediction, as illustrated in Figure~\ref{fig:Frame} (b). We use the last layer representations $(\boldsymbol{e}_p, \boldsymbol{e}_r)$ as neural abstractions for corresponding causal notions $(P, R)$, and identify an orthogonal low-rank matrix $\boldsymbol{O}$ that best matches the influence space of $R \to P$. By editing the preference representation in this space and converting the result back to the original space using $\boldsymbol{O}^T$, we can effectively predict user behavior, capturing both direct relevance and preference effects, as shown by the modified causal graph transitioning from the top left to the top right in the figure.
The disentanglement of relevance and preference further allows us to reconstruct a fine-grained modeling space, as depicted in Figure~\ref{fig:Frame} (c). 

\subsection{Preference Editing}
In this section, we first mathematically formulate the ideal behavior modeling method based on relevance and preference effects, then introduce a preference editing method to eliminate the relevance influence on effect to achieve this. 

Existing joint modeling frameworks typically combine the estimated effects $\hat{p}, \hat{r}$ as in Equation~\eqref{eq:fixed_scoring}. These frameworks assume that there are only direct effects of $P,R$ on $B$, i.e., only black arrows exist in the causal graph (Figure~\ref{fig:CauVenn} (a)). However, they overlook the relevance effect on preference (the red arrow), which can produce the indirect relevance effect on user behaviors (the path $R\to P \to B$ in the graph), potentially leading to biased outcomes, as highlighted by \citet{guo2020survey}. Therefore, the ideal modeling formula should be:
\begin{gather}
 \hat{p}_c = f(\hat{p})-g(\hat{r}) \label{eq:idea_cali}\\
    \hat{y} = \hat{r}^\delta\cdot\hat{p}_c \label{eq:idea_pred}
\end{gather}
where $\hat{p}_c$ represents the calibrated preference prediction, which mitigates the relevance effect in the initial prediction $\hat{p}$ as in Equation~\eqref{eq:idea_cali}, effectively removing the indirect relevance effect on behaviors. In Equation~\eqref{eq:idea_cali}, $f$ and $g$ are specifically designed functions to achieve the disentanglement purpose, culminating in a pure preference prediction in Equation~\eqref{eq:idea_pred}.


To find the optimal functions $f$ and $g$, which operate on the causal predictions $\hat{p}$ and $\hat{r}$, we can align these high-level causal signals with the low-level neural representations in deep models. This alignment allows us to alternatively determine the optimal transformations on neural representations in a learning-based manner~\cite{geiger2021causal}. Specifically, we treat the hidden representation from the final layer of preference models as the base representation and that from relevance models as the source representation. According to \citet{geiger2024finding}, a low-rank neural space can be identified that best matches the causal intervention space from the source effect to the base. Importantly, modifying the base representation within this low-rank space preserves information unrelated to the intervention. Building on this insight, we aim to learn the optimal orthogonal low-rank projection matrix as the desired transformation:
\begin{gather}
\boldsymbol{e}_{p_c} = \boldsymbol{O}^T(\boldsymbol{O}\boldsymbol{e}_p-\boldsymbol{O}\boldsymbol{e}_r) \label{eq:pref_edit}
\end{gather}
Let $\boldsymbol{e}_{p_c}$, $\boldsymbol{e}_p$, and $\boldsymbol{e}_r$ represent the final layer's hidden representations for $\hat{p}_c$, $\hat{p}$, and $\hat{r}$, respectively. We introduce $\boldsymbol{O} \in \mathbb{R}^{H \times D}$, a learnable low-rank projection matrix with orthogonal rows, where $H$ is the dimension of the last layer, and $D$ is the dimensionality of the low-rank subspace. Since $\boldsymbol{O}$ is orthogonal, $\boldsymbol{O}^T$ serves as its inverse, allowing the conversion of the edited representation back into the original space.
The calibrated preference can then be computed based on the edited preference representation: 
\begin{gather}
    \hat{p}_c = \boldsymbol{W}_p\boldsymbol{e}_{p_c}+\boldsymbol{b}_p \label{eq:pref_edit_pred}
\end{gather}
where $\boldsymbol{W}_p$ and $\boldsymbol{b}_p$ are the transformation matrix and bias for the linear decoder for preference prediction, respectively.


The process of preference editing is visualized in Figure~\ref{fig:Frame} (b). We begin by setting $\boldsymbol{e}_p$ and $\boldsymbol{e}_r$ as low-level representations for the high-level causal notions $P$ and $R$, respectively. Next, we learn an orthogonal space $\boldsymbol{O}$ that best aligns with the intervention space of $R \to P$. In the figure, $\boldsymbol{O} \in \mathbb{R}^{3 \times 2}$ is illustrated as an orthogonal projection matrix, with the corresponding space represented as a two-dimensional surface spanned by $\boldsymbol{O}_x$ and $\boldsymbol{O}_y$.
Within this intervention space, we subtract the relevance influence ($\boldsymbol{O} \boldsymbol{e}_r$, shown as the red dotted vector) from the preference representation ($\boldsymbol{O} \boldsymbol{e}_p$, the yellow dotted vector), obtaining the unbiased user preference (the blue dotted vector). We then project it back to the original behavior modeling space using $\boldsymbol{O}^T$, allowing us to decode it for unbiased preference prediction. This process culminates in the top-right causal graph, which directly models both preference and relevance based on behavioral signals and allows the segmentation of the modeling space as Figure~\ref{fig:CauVenn} (b).

\subsection{Adaptive Fusion}\label{sec:adap}
This section introduces a dual-level adaptive fusion method to guide the joint model toward learning from the appropriate modeling space, addressing Problem 2 discussed in Section~\ref{sec:pre_problems}.

From Equation~\eqref{eq:fixed_scoring} and Equation~\eqref{eq:idea_pred}, we observe that existing scoring methods will predict positive user behavior only when $\hat{p}_c(\hat{p}) \to 1$ and $\hat{r} \to 1$. Equation~\eqref{eq:idea_pred} can be rewritten as:
\begin{align}
     \hat{y} &= \hat{r}^\delta\cdot\hat{p}_c = \hat{r}^{\delta-1}\cdot\hat{p}_c\hat{r}\nonumber\\
     &= \hat{r}^{\delta-1}\cdot 
    \begin{pmatrix} 
    1&
    0
    \end{pmatrix}
    \begin{pmatrix} 
    \hat{p}_c\\
    1-\hat{p}_c
    \end{pmatrix}
    \begin{pmatrix} 
    \hat{r}&
    1-\hat{r}
    \end{pmatrix}
        \begin{pmatrix} 
    1\\
    0
    \end{pmatrix} \nonumber\\
    &=\hat{r}^{\delta-1}\cdot 
    \begin{pmatrix} 
    1&
    0
    \end{pmatrix}
    \begin{pmatrix} 
    \hat{p}_c\cdot\hat{r} & \hat{p}_c\cdot(1-\hat{r})\\
    (1-\hat{p}_c)\cdot\hat{r}&  (1-\hat{p}_c)\cdot (1-\hat{r}) 
    \end{pmatrix}
    \begin{pmatrix} 
    1\\
    0
    \end{pmatrix} \label{eq:Golb_init}
\end{align}
Denoting the $2\times 2$ matrix in Equation~\eqref{eq:Golb_init} as 
$
\begin{pmatrix} 
\mathbb{P}_{11} & \mathbb{P}_{10}\\
\mathbb{P}_{01} & \mathbb{P}_{00}
\end{pmatrix}
$,
we can find that $\mathbb{P}_{ij}$ is the prediction for the relevance-preference status $P = i, R = j$. For example, $\mathbb{P}_{10} = \hat{p}_c \cdot (1 - \hat{r})$ is the probability that this sample belongs to Area\#1 or Area\#2 where $P = 1, R = 0$ as listed in Table~\ref{tab:Area_num}. Notice that only $\mathbb{P}_{11}$ will influence the final prediction, as the probability for the remaining areas is rendered as zero by left multiplication with
$
\begin{pmatrix} 
1 & 0 
\end{pmatrix}
$
and right multiplication with
$
\begin{pmatrix} 
1 \\
0 
\end{pmatrix}.
$
The corresponding modeling space is visualized in Figure~\ref{fig:Frame} (c-\MakeUppercase{\romannumeral 1}), where only Area\#3 ($\mathbb{P}_{11}$) is properly learned. In this Venn diagram, Area\#1, 2 ($\mathbb{P}_{10}$) and Area\#4, 5 ($\mathbb{P}_{01}$) are hard to distinguish as they are merged with Area\#0 ($\mathbb{P}_{00}$).
To model the nuanced pattern of these areas, the global adaptive fusion replaces 
$
\begin{pmatrix} 
1 & 0 
\end{pmatrix}
$
and
$
\begin{pmatrix} 
1 \\
0 
\end{pmatrix}
$
in Equation~\eqref{eq:Golb_init} with learnable parameters:
\begin{align}
\hat{y}_g &=\hat{r}^{\delta-1}\cdot 
\boldsymbol{\alpha}
    \begin{pmatrix} 
 \mathbb{P}_{11} & \mathbb{P}_{10}\\
\mathbb{P}_{01}&\mathbb{P}_{00}
    \end{pmatrix}
\boldsymbol{\beta}^T\nonumber\\
&=\hat{r}^{\delta-1} (\mathbb{P}_{11}\alpha_1\beta_1+\mathbb{P}_{10}\alpha_1\beta_0+\mathbb{P}_{01}\alpha_0\beta_1+\mathbb{P}_{00}\alpha_0\beta_0)
\label{eq:Golb_score}
\end{align}
where $\boldsymbol{\alpha} = (\alpha_1, \alpha_0)$ and $\boldsymbol{\beta} = (\beta_1, \beta_0)$ are learnable parameters, and $\hat{y}_g$ stands for the behavior prediction with global adaptive fusion. Since $\sum_{ij} \mathbb{P}_{ij} = 1$ and $\boldsymbol{\alpha}, \boldsymbol{\beta}$ are usually non-zero, Equation~\eqref{eq:Golb_score} can distinguish the differences of different patterns when there is a clear result. For example, suppose $\mathbb{P}_{01} = 0.99$, the joint model is able to independently learn the distinct pattern for Area\#4 and Area\#5 (the area for $P = 0, R = 1$).

However, the global adaptive fusion is still insufficient to model the entire diagram. It divides the modeling space into four parts based on the relevance-preference status, but users exhibit varying sensitivities to relevance, depending on different queries and items. This results in different behavior signals for the same relevance-preference status~\cite{hong2024print}. As illustrated in Figure~\ref{fig:Frame} (c-\MakeUppercase{\romannumeral 2}), the global adaptive fusion fails to distinguish between Area\#1 and Area\#2, as well as between Area\#4 and Area\#5. To address this issue, we further propose the local adaptive fusion:
\begin{gather}
    \hat y_l=\hat y_g\cdot\mathcal{F}(\boldsymbol{u},\boldsymbol{v},\boldsymbol{q};\hat y_g,y)\label{eq:final_pred}
\end{gather}
where $\hat{y}_g$ is the global adaptive prediction from Equation~\eqref{eq:Golb_score}, and $y$ is the ground truth of user behaviors (e.g., $y = 1$ denotes `Click' for click modeling). $\mathcal{F}$ aims to fit the gap between the global score $\hat{y}_g$ and the heterogeneous behavior signals $y$ of similar relevance-preference statuses, ultimately achieving nuanced modeling as illustrated in Figure~\ref{fig:Frame} (c-\MakeUppercase{\romannumeral 3}).
Note that only $\boldsymbol{u}, \boldsymbol{v}, \boldsymbol{q}$ are fed to $\mathcal{F}$ during the feed-forward loop, while $\hat{y}_g$ and $y$ are used to construct the training objective for $\mathcal{F}$ in the training loop. This fills the gap between the diverse $y$ and the similar global predictions $\hat{y}_g$ in Area\#1\&2 and Area\#4\&5, enabling end-to-end training and serving.

\subsection{Inference \& Optimization}
We illustrate the inference and optimization process of \name in this section. Overall, \name operates on the scoring stage, where the preference model and the relevance model remain unaffected. 

For inference, the relevance is the same as calculated in Equation~\eqref{eq:RM}. We replace the last layer representation of the preference model as in Equation~\eqref{eq:pref_edit} and predict the preference effect following Equation~\eqref{eq:pref_edit_pred}. Finally, we propose the dual-level adaptive fusion, merging two effects as in Equation~\eqref{eq:final_pred}. 

With the disentangled relevance and preference effects and the well-segmented modeling space, \name is able to learn fined-grained patterns without the human-labeled relevance signals in an end-to-end manner. As there are no auxiliary losses for \name, we can optimize the framework with the mere behavior modeling loss and apply the gradient-descent strategy. Take the click modeling task (click-through rate prediction) as an example, we can train \name with binary-cross entropy loss:
\begin{gather}
    \mathcal{L} =  \sum_{y\in \mathcal{T}}y\cdot \log \hat y+\left(1-y\right) \cdot \log \left(1-\hat y\right)
\end{gather}
where we use $\mathcal{T}$ to denote the training set.

\begin{table}[t]
\tabcolsep=0.12cm 
\centering
\caption{Dataset statistics.}
 \vspace{-2mm}
\label{tab:dataset}
\begin{tabular}{@{}ccccccc@{}}

\toprule
       & \# User & \# Item    & \# Query & \# Interaction & \# Session \\ \midrule
KuaiSAR   & 19,851 & 1,974,165     & 126,027     & 3,038,362    & 186,268      \\
JDSearch &  - & 8,305,606 &  110,763  & 15,439,583    & 173,825\\
Private & 9,426 & 196,645 &  123,941  &    45,245,013   & 1,165,596 \\
\bottomrule
\end{tabular}%
\vspace{-3mm}
\end{table}

\begin{table*}[t]
\centering
\tabcolsep=0.1cm 
\caption{Overall performance comparison. }
\label{tab:OV}
\resizebox{\textwidth}{!}{
\begin{tabular}{c|c|c|cccc|cccc|cccc}
\toprule
\multirow{2}{*}{PM} & \multirow{2}{*}{RM} & \multirow{2}{*}{JM} & \multicolumn{4}{c}{Private} & \multicolumn{4}{|c}{KuaiSAR} & \multicolumn{4}{|c}{JDSearch} \\ \cline{4-15} 
                    &                     &                     & AUC  & LogLoss  & NDCG & HR & AUC  & LogLoss  & NDCG & HR & AUC  & LogLoss  & NDCG  & HR \\ \hline
\multirow{3}{*}{-}  & DSSM                & \multirow{3}{*}{-}  &  0.6468& 0.05674&0.1628&0.2867   &    0.5687&0.3662&0.4088&0.7085   & 0.6554&0.09837&0.1916&0.3356 \\
                      & QEM                  &      & 0.6491&0.05606&0.1635&0.2881 & 0.5669&0.3640&0.4087&0.7102  & 0.6564 &0.09712&0.1931&0.3355  \\
                      & HEM                  &      & 0.6684&0.05577&0.1634&0.2883&0.6146&0.3561&0.4091& 0.7100  & 0.6560& 0.09705&0.1984&0.3426    \\ \hline
\multirow{18}{*}{MLP} & -                 & -    & 0.6686&0.05577&0.1632&0.2879&0.6123&0.3566&0.4098&0.7107  &  0.6557& 0.09705&0.1961& 0.3394  \\\cline{2-15} 
                      & \multirow{5}{*}{DSSM}               & Base   &0.6690&0.05572&\underline{0.1649}&0.2904&0.6206&0.3556& 0.4112&0.7117 &0.6669& \textbf{0.09670}& 0.1968& 0.3445 \\ 
                      &            & CLK    &0.6692& 0.05571&0.1648& 0.2903&0.6205& 0.3556& 0.4110& 0.7115 &0.6657&0.09682& 0.1960& 0.3427\\
                        &            & NISE    &0.6689& 0.05571&0.1648& \underline{0.2906}&\underline{0.6207}&0.3555& \underline{0.4113}&\underline{0.7119} &  0.6669& \textbf{0.09670}& 0.1969&0.3444         \\
                      &            & PRINT   &\underline{0.6696}&\underline{0.05564}&0.1644 &0.2897&0.6195& \underline{0.3552}& 0.4084& 0.7100 &\underline{0.6696}&0.09724&\underline{0.2018}& \underline{0.3477}\\
                      &                 & \name    &\textbf{0.6710}*& \textbf{0.05559}*&\textbf{0.1660}*& \textbf{0.2920}*&\textbf{0.6229}*& \textbf{0.3549}&\textbf{0.4168}*& \textbf{0.7145}* & \textbf{0.6824}*& 0.09685&\textbf{0.2052}*& \textbf{0.3620}* \\ \cline{2-15} 
& \multirow{6}{*}{QEM} & Base    & 0.6696&0.05566&0.1642&0.2897&0.6206& 0.3554&0.4112& 0.7118& 0.6783&\underline{0.09874}&0.1979&0.3550 \\
                      &                      & CLK  &\underline{0.6701}&\underline{0.05565}&\underline{0.1662}&\underline{0.2922}&0.6205& 0.3553&0.4113&0.7112 &0.6787&0.09878&0.1960&0.3524  \\
                      &                      & NISE &0.6698&0.05570&0.1657&0.2913&0.6207 &0.3553&\underline{0.4115}& \underline{0.7122}  &  0.6793& 0.09882&0.1963&0.3539  \\
                      &                      & DCMT &0.6699&0.05571&0.1656&0.2913&\underline{0.6214}&\textbf{0.3550}& 0.4110&\underline{0.7122}& \underline{0.6796}&0.09888&0.1972&0.3537  \\
                      &                      & PRINT & 0.6689& 0.05568&0.1638& 0.2889&0.6208&0.3556&0.4095&0.7110&0.6707&0.09930 &\underline{0.1987}& \underline{0.3553} \\
                      &                      &  \name    & \textbf{0.6707}*&\textbf{0.05561}*&\textbf{0.1663} &\textbf{0.2923} &\textbf{0.6227}*&\underline{0.3551}&\textbf{0.4143}*& \textbf{0.7132}* & \textbf{0.6904}*&\textbf{0.09862}*&\textbf{0.2053}*&\textbf{0.3635}*  \\ \cline{2-15} 
& \multirow{6}{*}{HEM} & Base    &0.6695&0.05570&0.1649&0.2903& \underline{0.6218}& 0.3553& \underline{0.4118}&\underline{0.7117}& 0.6774& 0.09884&\underline{0.2032}&\underline{0.3616} \\
                      &                      & CLK  &0.6694&\underline{0.05568}& 0.1652& 0.2905& 0.6217& 0.3553&0.4115&0.7114& 0.6770&\underline{0.09863}&0.1989&0.3546 \\
                      &                      & NISE &0.6695&\underline{0.05568}&0.1647& 0.2901&\underline{0.6218}& 0.3553&0.4116& \underline{0.7117} &0.6776&0.09886&0.2029&0.3609\\
                      &                      & DCMT & \underline{0.6699}&0.05570 & \underline{0.1659}&\underline{0.2912}&0.6214&\underline{0.3552}&0.4101& 0.7116  & \underline{0.6778}& 0.09895&0.1998& 0.3574 \\
                      &                      & PRINT &0.6693& \underline{0.05568}&0.1645& 0.2900&0.6209& 0.3554&0.4094&0.7109& 0.6758&0.09902&0.1956 &0.3520\\ 
                      &                      & \name &\textbf{0.6702}& \textbf{0.05563}* &\textbf{0.1663}&\textbf{0.2924}*&\textbf{0.6230}*& \textbf{0.3551}&\textbf{0.4138}*&\textbf{0.7122} & \textbf{0.6876}*&\textbf{0.09814}*&\textbf{0.2085}*&\textbf{0.3690}* \\ \hline
\multirow{18}{*}{DCN} & -                 & -    &0.6603&0.05598&0.1607 &0.2839& 0.6183& 0.3564&0.4144& 0.7126& 0.6738&0.09806&0.2096& 0.3634  \\\cline{2-15} 
& \multirow{5}{*}{DSSM}                & Base    &0.6608&0.05595&0.1609 &0.2845 &\underline{0.6184}&\underline{0.3561}&0.4132 &0.7120& 0.6702&\underline{0.09660} &0.2022 &0.3504  \\ 
                      &            & CLK    &\underline{0.6615}&0.05595 &\underline{0.1615}&\underline{0.2852}&0.6183&\underline{0.3561}&0.4131 &0.7117 &0.6686& 0.09674& 0.2009 &0.3481\\
                        &            & NISE    &0.6612&\underline{0.05590}&0.1613&\underline{0.2852}&\underline{0.6184}&\underline{0.3561}&\underline{0.4133}&\underline{0.7124} &0.6702& \underline{0.09660} &0.2022& 0.3503           \\
                      &            & PRINT   &0.6599& 0.05605&0.1598&0.2830&0.6165 &\underline{0.3561}&0.4108& 0.7103 &\underline{0.6724}& 0.09712&\underline{0.2108}&\underline{0.3629}\\
                      &            & \name    &\textbf{0.6666}*& \textbf{0.05577}*&\textbf{0.1643}*& \textbf{0.2891}*&\textbf{0.6200}*& \textbf{0.3557}* &\textbf{0.4145}*& \textbf{0.7133} & \textbf{0.6915}*& \textbf{0.09654}&\textbf{0.2125}& \textbf{0.3702}* \\\cline{2-15} 
& \multirow{6}{*}{QEM} & Base   &\underline{0.6606}&0.05604&0.1610&\underline{0.2845}& 0.6192& 0.3555& 0.4118& 0.7107&  0.6874&0.09894&0.2143&0.3745  \\
                      &                      & CLK  &0.6604&0.05611&\underline{0.1612} &0.2843&0.6190& 0.3555&0.4118&0.7097 &  0.6857&0.09931&0.2120&0.3706 \\
                      &                      & NISE &0.6603&0.05611&0.1611&0.2843&0.6193& 0.3555&0.4119&0.7106 & \underline{0.6876}&0.09887&0.2142&0.3749  \\
                      &                      & DCMT & 0.6602&0.05601&0.1604&0.2836&\underline{0.6208}& \underline{0.3553}&0.4134& 0.7120  &  0.6843&0.09907&\textbf{0.2166}&\textbf{0.3776} \\
                      &                      & PRINT &\underline{0.6606}& \underline{0.05592}&0.1604& 0.2838&0.6202& 0.3555& \underline{0.4138}& \underline{0.7125}&0.6842& \textbf{0.09877}&0.2136& 0.3736 \\
                      &                      & \name & \textbf{0.6663}*&\textbf{0.05566}*&\textbf{0.1632}*&\textbf{0.2873}* &\textbf{0.6243}*& \textbf{0.3547}*&\textbf{0.4163}*&\textbf{0.7142}*  & \textbf{0.6942}*& \underline{0.09882}&\underline{0.2154}&\underline{0.3767}  \\ \cline{2-15} 
& \multirow{6}{*}{HEM} & Base    &\underline{0.6618}&\underline{0.05590}&0.1607&0.2836&0.6229& \underline{0.3548}&\underline{0.4132}&\underline{0.7133}&  0.6852&0.09900&\underline{0.2184}&0.3798  \\
                      &                      & CLK  &0.6610&0.05595&0.1602&0.2830&\underline{0.6231}& \underline{0.3548}&0.4131&0.7124& 0.6868 &0.09912&0.2167&0.3766 \\
                      &                      & NISE &0.6611&0.05595&0.1602 &0.2834&0.6229& \underline{0.3548}&0.4131&0.7129 &  0.6855& 0.09898&0.2178& 0.3792  \\
                      &                      & DCMT &0.6603&0.05608&\underline{0.1612} &\underline{0.2841} & 0.6216& 0.3550&\underline{0.4132}&0.7122  &  \underline{0.6874}&0.09890&  \underline{0.2184}& \underline{0.3804}  \\
                      &                      & PRINT &0.6603& 0.05599&0.1601& \underline{0.2841}&0.6209&0.3551&0.4119&0.7112&0.6841& \underline{0.09875} &0.2142&0.3752 \\
                      &                      & \name  &\textbf{0.6678}*&\textbf{0.05562}*&\textbf{0.1647}* &\textbf{0.2896}* &\textbf{0.6236}& \textbf{0.3545}&\textbf{0.4152}*&\textbf{0.7144}*  & \textbf{0.6938}*& \textbf{0.09871}&\textbf{0.2194}&\textbf{0.3855}*  \\ \bottomrule
\end{tabular}
}
\\``\textbf{{\Large *}}'' indicates the statistically significant improvements. (i.e., one-sided t-test with $p<0.05$) over the runner-up method.
\vspace{-3mm}
\end{table*}

\section{Experiment}
We present experimental settings and extensive empirical results in this section. Additional experiments are provided in \textbf{Appendix~\ref{appsec:res}}.

\subsection{Experiment Setting}
\subsubsection{\textbf{Dataset}}
\name is evaluated on two public datasets and data collected on our private e-commerce platform. 
\textbf{KuaiSAR}~\cite{sun2023kuaisar}
, is a unified search and recommendation dataset released by Kuaishou, a short-video platform. We split out the video search log to test the effectiveness of \name in scenarios other than e-commerce.
\textbf{JDSearch}~\cite{liu2023jdsearch}
, constructed by JD.com, a popular online shopping platform. This is a typical e-commerce scenario.
\textbf{Private}, collected from the one-month daily log on our private e-commerce platform. 
It is notable that all three datasets contain \textit{real} queries (anonymized for public two). 
The detailed statistics, including the number of users, items, queries, interactions, and search sessions, are listed in Table~\ref{tab:dataset}.
For data splitting, 
search sessions in the first 80\% time period are the training set, the middle 10\% is the validation, and the remaining 10\% is the test set.
    
\vspace{-2mm}
\subsubsection{\textbf{Backbones \& Baselines}} 
To verify the flexibility of our model, we integrate various relevance and preference backbone models for comparison. 
\textbf{Relevance Models}: DSSM~\cite{huang2013learning} (ad-hoc), QEM~\cite{ai2017learningHEM} (ad-hoc) and HEM~\cite{ai2017learningHEM} (personalized). 
\textbf{Preference Models}: MLP~\cite{zhang2016ffn} and DCN~\cite{wang2017dcn}. 
Besides, we compare several \textbf{Joint Modeling Baselines}: 
CLK~\cite{yao2021learningpositionbias}, 
NISE~\cite{huang2024utilizing}, DCMT~\cite{zhu2023dcmt}, 
and PRINT~\cite{hong2024print}.

\vspace{-2mm}
\subsubsection{\textbf{Evaluation Metrics}}

Following previous works~\cite{guo2017deepfm}, we adopt Area Under the ROC Curve, denoted as \textbf{AUC} ($\uparrow$), and \textbf{LogLoss} ($\downarrow$) as the evaluation metrics. 
Besides, to show the performance in ranking, we also consider Top-10 Hit Rate, \ie \textbf{HR} ($\uparrow$) and Top-10 Normalized Discounted Cumulative Gain, \ie \textbf{NDCG} ($\uparrow$) as metrics. It is worth noting that we only take the sessions with positive samples into account when calculating these two metrics.
\vspace{-2mm}
\subsubsection{\textbf{Implementation Details}}
For the basic joint modeling structure, we employ M3E~\cite{Moka} to encode textual contents in Private. We encode the anonymous text within JDSearch and KuaiSAR, as well as user behavior sequences in all datasets following UniSAR~\cite{shi2024unisar}. Other features are all encoded as one-hot signals and converted to embeddings. Above this, we obtain 64-dimensional $\boldsymbol{q},\boldsymbol{v},\boldsymbol{u}$ and feed to the relevance and preference model. Units of the prediction layer, i.e., the last MLP component of $\mathrm{RM},\mathrm{PM}$, is $[64,32,1]$, indicating $H=32$ for $\boldsymbol{e}_p,\boldsymbol{e}_{p_c},\boldsymbol{e}_r$. The low-rank projection space is 16-dimensional, i.e., $D=16$ for $\boldsymbol{R}$. For global adaptive fusion, the initial state is $(1,0.5)$ for $\boldsymbol{\alpha}$ and $\boldsymbol{\beta}$.

\subsection{Overall Performance}
This section comprehensively assesses \name on various relevance and preference backbones, comparing it with existing methods crafted to enhance the joint modeling. Each experiment is repeated ten times with different seeds for reliable results. We present the mean of these ten trials with four valid decimal places, and implement the one-tailed t-test to confirm the superiority of \textbf{the best-performing method} (in bold) over \underline{the runner-up} (underlined). 
The result is presented in Table~\ref{tab:OV}, where `PM' denotes the preference backbone, `RM' is the relevance model, `JM' stands for the method to boost joint modeling frameworks, and `-' means no solution is applied. For example, the first three rows that with `PM' and `JM' as `-' are relevance-only methods. 

Overall, the joint modeling approach outperforms methods that consider only relevance or preference due to its integration of both factors. This superiority underscores the importance of developing joint modeling frameworks. Additionally, we observe that the personalized relevance model and the preference model exhibit similar performance levels (for example, MLP versus HEM on Private and JDSearch). Both models take $\boldsymbol{u},\boldsymbol{v},\boldsymbol{q}$ as inputs and are trained on the same behavioral signals. They are likely to present similar results despite architectural differences. This observation suggests that relying solely on behavioral signals is insufficient for models to learn the single intended effect, as the two effects are intertwined. This highlights the necessity for disentangled effects.
    
When comparing `CLK' with the `Base' of joint modeling frameworks, it's not always advantageous to learn relevance from behavior data, as suggested by \cite{yao2021learningpositionbias}. While the original literature indicates that relevance models can benefit from click data, this may not hold in the context of behavior modeling. This creates a modeling loop: `behavior -> relevance -> behavior', which suffers from a lack of human-labeled relevance data to refine the relevance and prevent overfitting on `behavior'. This situation calls for a design that does not depend on human-labeled data.
    
For CVR solutions, NISE, and DCMT, although they partly address the disentangled effect and the misaligned modeling space, they still fail to significantly enhance the joint modeling framework (`Base'). Because the click signal is observed in CVR problems, while both relevance and preference signals are unavailable in our scenario. This necessitates a modeling space reconstruction solution specifically tailored to our joint behavior modeling challenges.
    
The existing adaptive fusion method, PRINT, fails to consistently improve joint modeling performance. It divides the relevance effect into three segments: positive, zero, and negative effects, corresponding to Area\#4\&5, Area\#0\&3, and Area\#1\&2, respectively. Thus, it struggles to discern subtle patterns within combined areas (e.g., Area\#1 in Area\#1\&2), leading to suboptimal outcomes. This calls for a more fine-grained adaptive fusion framework that provides stable predictions across areas.
    
The proposed \name outperforms others by incorporating preference editing and adaptive fusion, effectively addressing the issue of entangled effects and reconstructing the correct modeling space. Our work delves into the nuances of the joint modeling problem and avoids the need for human-labeled relevance data.


\begin{table}[t]
\centering
\caption{Ablation study on JDSearch with MLP.}
\label{tab:abl_JD}
\begin{tabular}{@{}ccccccc@{}}
\toprule
 & \multicolumn{2}{c}{DSSM} & \multicolumn{2}{c}{QEM} & \multicolumn{2}{c}{HEM} \\ 
 & AUC        & NDCG        & AUC        & NDCG       & AUC        & NDCG       \\\midrule
 Base   &     0.6669 &     0.1968        &      0.6783 &   0.1979         & 0.6774    & 0.2032   \\        \name  &   0.6824 & 0.2052  & 0.6904  &  0.2053     &  0.6876 &  0.2085  \\
\name-2 &       0.6671 & 0.1969       &    0.6808  &0.2021  &  0.6796 & 0.1990    \\
\name-3 &  0.6673&0.1984      &  0.6821   &  0.2040    &  0.6818    & 0.2050           \\
\name-4 &    0.6683&0.1979                 &    0.6832 &  0.1975     &  0.6842   &0.2070  \\
\name-5 &   0.6813&0.2045&0.6876&0.2047&0.6851&0.2060
\\ \bottomrule
\end{tabular}
\end{table}

\subsection{Ablation Study}
We investigate core components of \name in this section, focusing on four designed variants to study the effects of orthogonal preference editing and dual-level adaptive scoring.
\begin{itemize}[leftmargin=*]
    \item \textbf{\name-1} drops the orthogonal constraint on $\boldsymbol{O}$. Specifically, it replaces Equation~\eqref{eq:pref_edit} with $\boldsymbol{e}_{p_c} = \boldsymbol{W}_2^T(\boldsymbol{W}_1\boldsymbol{e}_p - \boldsymbol{W}_1\boldsymbol{e}_r)$, where $\boldsymbol{W}_1$ and $\boldsymbol{W}_2$ are learnable matrices with the same dimensions as $\boldsymbol{O}$.
\item \textbf{\name-2} omits the adaptive fusion component to isolate the effectiveness of the remaining component, i.e., preference editing.
\item \textbf{\name-3} excludes global adaptive fusion to assess its impact, i.e., $\hat y_g$ in Equation~\eqref{eq:final_pred} is replaced by $\hat y$ from Equation~\eqref{eq:pref_edit_pred}.
\item \textbf{\name-4} skips Equation~\eqref{eq:final_pred} to evaluate its efficacy, focusing on the effectiveness of local adaptive scoring. 
\item \textbf{\name-5} omits the preference editing module and directly constructs the modeling space using the impure preference effect.
\end{itemize}

We evaluate these variants on JDSearch with MLP. The result is listed in Table~\ref{tab:abl_JD}, where AUC and NDCG are presented. 
From this table, we can find the original design achieves the best results, outperforming all variants. This suggests that the combination of adaptive fusion with disentangled effects greatly enhances effectiveness, highlighting the success of the reconstructed modeling space by \name.

\begin{figure}[!t]
\centering
\includegraphics[width=0.99\linewidth]{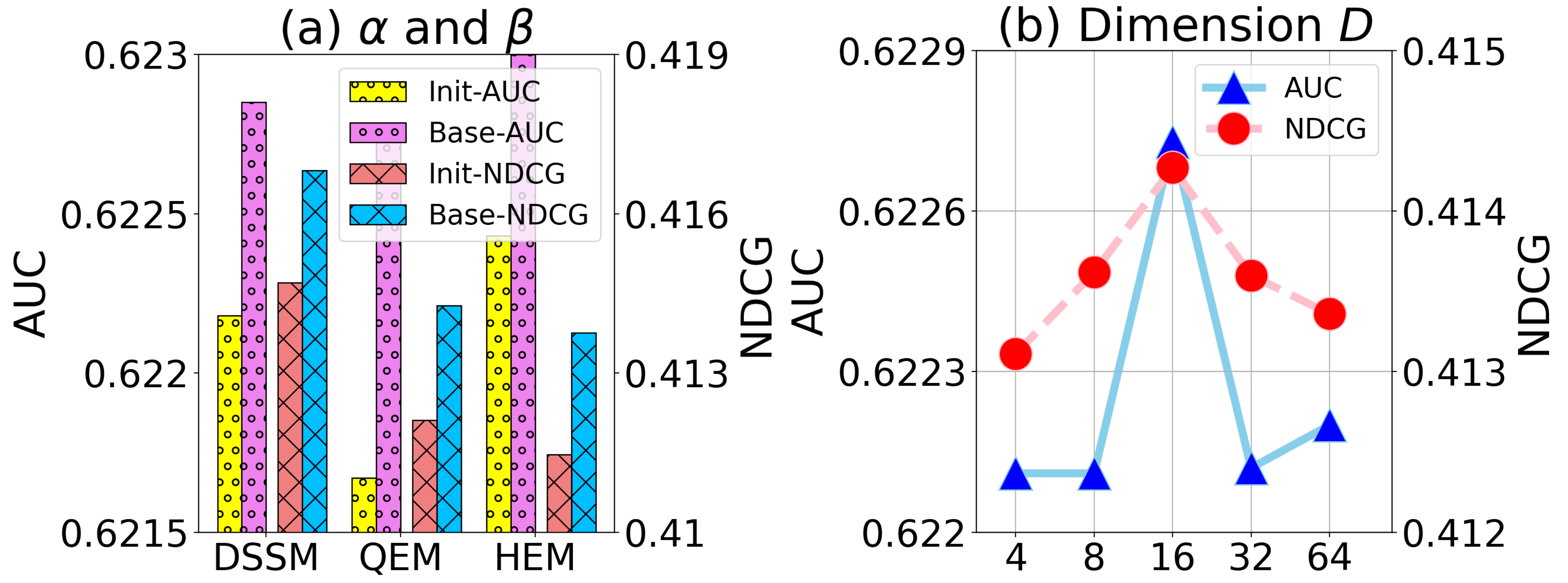}
\caption{The hyper-parameter experiments on KuaiSAR.}
\label{fig:exp_hyper}
\vspace{-4mm}
\end{figure}

\subsection{Parameter Analysis}
This section explores the sensitivity of \name to key parameters, specifically the dimension $D$ of the low-rank projection space and the initial values of $\boldsymbol{\alpha}$ and $\boldsymbol{\beta}$ for the global adaptive fusion.
We varied $D$ across the set $[4,8,16,32,64]$ and tested initial values of $\boldsymbol{\alpha}=\boldsymbol{\beta}=(1,0.5)$, denoted as \textit{Base} in the Figure~\ref{fig:exp_hyper} (a), and $\boldsymbol{\alpha}=\boldsymbol{\beta}=(0.5,0.5)$, denoted as \textit{Init}. We change one parameter at a time and evaluate \name's performance on KuaiSAR using MLP. 

From \textbf{Figure~\ref{fig:exp_hyper} (a)}, the distinct initialization of $\boldsymbol{\alpha}$ and $\boldsymbol{\beta}$ as $(1,0.5)$ is beneficial. This initialization appears to provide \name with the prior knowledge that positive relevance or preference usually leads to positive behaviors.
From \textbf{Figure~\ref{fig:exp_hyper} (b)}, the performance shows an inverse V-shaped pattern as $D$ increases. A lower-dimensional space may be insufficient to capture the influence of relevance on preference, explaining the initial performance boost as $D$ grows. However, an overparameterized space might introduce irrelevant information. causing the performance drop. 

\subsection{Model Visualization}
\begin{figure}[t]

    \centering
    \includegraphics[width=0.55\linewidth]{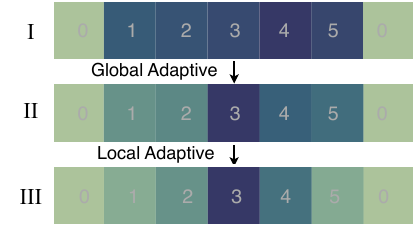}
    \caption{Model visualization of adaptive fusion.}
   \vspace{-5mm}
    \label{fig:vis}
\end{figure}

This section visualizes the effects of dual-level adaptive fusion described in Section~\ref{sec:adap}. 
We load the checkpoint of \name with MLP and HEM on KuaiSAR, subsequently calculating the average behavior predictions for each designated area (Area\#0-5), illustrated via a heatmap in Figure~\ref{fig:vis}. 
We classify samples with the top 20\% relevance scores as relevant and the same for preference to construct areas.
For illustrative clarity, we incorporate an additional Area\#0 following Area\#6. The modeled areas depicted in Figure~\ref{fig:vis} (\MakeUppercase{\romannumeral 1}-\MakeUppercase{\romannumeral 3}) correspond directly to the configuration shown in Figure~\ref{fig:Frame}(c).
From this figure, several observations emerge: (i) The fixed fusion in Equation~\ref{eq:pref_edit_pred} fails to differentiate Area\#2\&4 and Area\#1\&5, collapsed as Area\#3. 
(ii) Through global adaptive fusion, \name successfully discriminates among four distinct areas: Area\#0, Area\#1\&2, Area\#3, and Area\#4\&5 (noted by similar color in the figure). (iii) The implementation of local adaptive fusion ultimately enables discerning fine-grained patterns across various areas. These findings are in accord with our theoretical analysis in Section~\ref{sec:adap}.
\section{Related Works}

E-commerce search models, also known as product search models, can be divided into three categories: ad-hoc search models, personalized product search, and the development of relevance models. Ad-hoc models~\cite{robertson2009probabilistic,huang2013learning,ai2017learningHEM} conceptualize e-commerce search as a user-independent relevance modeling problem. While personalized search endeavors to synthesize user behavior sequences to generate personalized results, which is enhanced through the utilization of knowledge graphs~\cite{ai2021modelHGN} and transformers~\cite{ai2019zeroZAM,bi2020transformerTEM}. \citet{xiao2019weakly} was the first to articulate the notion of relevance modeling, which has inspired numerous subsequent efforts aimed at designing pre-trained relevance models to be integrated within a preference modeling framework, operating as a joint behavior modeling approach. Techniques such as knowledge distillation~\cite{jiang2020bert2dnn}, self-supervised learning~\cite{chen2023beyond}, and pretraining pipelines~\cite{zan2023spm} are employed in this context. Nonetheless, these frameworks currently exhibit a heavy reliance on laboriously collected relevance data; although some enhancement methods based on click data have been proposed~\cite{yao2021learningpositionbias}, there remains unexplored for the joint modeling framework to address latent issues or augment its applicability. A relevant contribution might be PRINT~\cite{hong2024print}, which personally merges the relevance effect during the fusion phase in a vague manner. 
Our study examines the challenges posed by the collapsed modeling space problem within the joint framework, and we introduce \name, which employs fine-grained, explicit adaptive fusion for easy deployment across diverse systems without relevance data.
Another relevant research area is the solution for the collapsed modeling space. Despite substantial attention from CVR prediction~\cite{zhu2023dcmt,huang2024utilizing,wang2022escm2} and impression bias~\cite{lin2024recrec,lee2022bilateral}, joint modeling strategies remain limited explorations. We are pioneering a thorough analysis of the collapsed modeling space within the context of relevance-preference joint modeling. 

\section{Conclusion}
This paper introduced \name, a novel joint modeling framework for e-commerce search systems. We begin with a comprehensive review of the existing frameworks and point out the collapsed modeling space problem. \name innovatively edits the preference representation to exclude the relevance influence on it, thus reconstructing the desired behavior modeling space with only direct effects from the preference and relevance. It then captures nuanced patterns within the fine-grained modeling space by dual-level adaptive fusion strategy. With these components, \name could be applied with various preference and relevance backbones without laborious collected relevance data. Our extensive experiments have supported the superiority of \name in search scenarios.

\begin{acks}
This research was partially supported by Research Impact Fund (No.R1015-23), Collaborative Research Fund (No.C1043-24GF), APRC - CityU New Research Initiatives (No.9610565, Start-up Grant for New Faculty of CityU), Hong Kong ITC Innovation and Technology Fund Midstream Research Programme for Universities Project (No.ITS/034/22MS), Huawei (Huawei Innovation Research Program), Tencent (CCF-Tencent Open Fund, Tencent Rhino-Bird Focused Research Program), Ant Group (CCF-Ant Research Fund), Alibaba (CCF-Alimama Tech Kangaroo Fund No. 2024002), and Kuaishou.
\end{acks}


\clearpage
\bibliographystyle{ACM-Reference-Format}
\bibliography{main}


\begin{thebibliography}{63}


\ifx \showCODEN    \undefined \def \showCODEN     #1{\unskip}     \fi
\ifx \showISBNx    \undefined \def \showISBNx     #1{\unskip}     \fi
\ifx \showISBNxiii \undefined \def \showISBNxiii  #1{\unskip}     \fi
\ifx \showISSN     \undefined \def \showISSN      #1{\unskip}     \fi
\ifx \showLCCN     \undefined \def \showLCCN      #1{\unskip}     \fi
\ifx \shownote     \undefined \def \shownote      #1{#1}          \fi
\ifx \showarticletitle \undefined \def \showarticletitle #1{#1}   \fi
\ifx \showURL      \undefined \def \showURL       {\relax}        \fi
\providecommand\bibfield[2]{#2}
\providecommand\bibinfo[2]{#2}
\providecommand\natexlab[1]{#1}
\providecommand\showeprint[2][]{arXiv:#2}

\bibitem[Ai et~al\mbox{.}(2019a)]%
        {ai2019zeroZAM}
\bibfield{author}{\bibinfo{person}{Qingyao Ai}, \bibinfo{person}{Daniel~N Hill}, \bibinfo{person}{SVN Vishwanathan}, {and} \bibinfo{person}{W~Bruce Croft}.} \bibinfo{year}{2019}\natexlab{a}.
\newblock \showarticletitle{A zero attention model for personalized product search}. In \bibinfo{booktitle}{\emph{Proceedings of the 28th ACM International Conference on Information and Knowledge Management}}. \bibinfo{pages}{379--388}.
\newblock


\bibitem[Ai and Narayanan.~R(2021)]%
        {ai2021modelHGN}
\bibfield{author}{\bibinfo{person}{Qingyao Ai} {and} \bibinfo{person}{Lakshmi Narayanan.~R}.} \bibinfo{year}{2021}\natexlab{}.
\newblock \showarticletitle{Model-agnostic vs. model-intrinsic interpretability for explainable product search}. In \bibinfo{booktitle}{\emph{Proceedings of the 30th ACM International Conference on Information \& Knowledge Management}}. \bibinfo{pages}{5--15}.
\newblock


\bibitem[Ai et~al\mbox{.}(2017)]%
        {ai2017learningHEM}
\bibfield{author}{\bibinfo{person}{Qingyao Ai}, \bibinfo{person}{Yongfeng Zhang}, \bibinfo{person}{Keping Bi}, \bibinfo{person}{Xu Chen}, {and} \bibinfo{person}{W~Bruce Croft}.} \bibinfo{year}{2017}\natexlab{}.
\newblock \showarticletitle{Learning a hierarchical embedding model for personalized product search}. In \bibinfo{booktitle}{\emph{Proceedings of the 40th International ACM SIGIR Conference on Research and Development in Information Retrieval}}. \bibinfo{pages}{645--654}.
\newblock


\bibitem[Ai et~al\mbox{.}(2019b)]%
        {ai2019explainableDREM}
\bibfield{author}{\bibinfo{person}{Qingyao Ai}, \bibinfo{person}{Yongfeng Zhang}, \bibinfo{person}{Keping Bi}, {and} \bibinfo{person}{W~Bruce Croft}.} \bibinfo{year}{2019}\natexlab{b}.
\newblock \showarticletitle{Explainable product search with a dynamic relation embedding model}.
\newblock \bibinfo{journal}{\emph{ACM Transactions on Information Systems (TOIS)}} \bibinfo{volume}{38}, \bibinfo{number}{1} (\bibinfo{year}{2019}), \bibinfo{pages}{1--29}.
\newblock


\bibitem[Bi et~al\mbox{.}(2020)]%
        {bi2020transformerTEM}
\bibfield{author}{\bibinfo{person}{Keping Bi}, \bibinfo{person}{Qingyao Ai}, {and} \bibinfo{person}{W~Bruce Croft}.} \bibinfo{year}{2020}\natexlab{}.
\newblock \showarticletitle{A transformer-based embedding model for personalized product search}. In \bibinfo{booktitle}{\emph{Proceedings of the 43rd International ACM SIGIR Conference on Research and Development in Information Retrieval}}. \bibinfo{pages}{1521--1524}.
\newblock


\bibitem[Bi et~al\mbox{.}(2021)]%
        {bi2021learningREM}
\bibfield{author}{\bibinfo{person}{Keping Bi}, \bibinfo{person}{Qingyao Ai}, {and} \bibinfo{person}{W~Bruce Croft}.} \bibinfo{year}{2021}\natexlab{}.
\newblock \showarticletitle{Learning a fine-grained review-based transformer model for personalized product search}. In \bibinfo{booktitle}{\emph{Proceedings of the 44th International ACM SIGIR Conference on Research and Development in Information Retrieval}}. \bibinfo{pages}{123--132}.
\newblock


\bibitem[Carmel et~al\mbox{.}(2020)]%
        {carmel2020irrlevant-amazon}
\bibfield{author}{\bibinfo{person}{David Carmel}, \bibinfo{person}{Elad Haramaty}, \bibinfo{person}{Arnon Lazerson}, \bibinfo{person}{Liane Lewin-Eytan}, {and} \bibinfo{person}{Yoelle Maarek}.} \bibinfo{year}{2020}\natexlab{}.
\newblock \showarticletitle{Why do people buy seemingly irrelevant items in voice product search? On the relation between product relevance and customer satisfaction in ecommerce}. In \bibinfo{booktitle}{\emph{Proceedings of the 13th international conference on web search and data mining}}. \bibinfo{pages}{79--87}.
\newblock


\bibitem[Chen et~al\mbox{.}(2023)]%
        {chen2023beyond}
\bibfield{author}{\bibinfo{person}{Zeyuan Chen}, \bibinfo{person}{Wei Chen}, \bibinfo{person}{Jia Xu}, \bibinfo{person}{Zhongyi Liu}, {and} \bibinfo{person}{Wei Zhang}.} \bibinfo{year}{2023}\natexlab{}.
\newblock \showarticletitle{Beyond Semantics: Learning a Behavior Augmented Relevance Model with Self-supervised Learning}. In \bibinfo{booktitle}{\emph{Proceedings of the 32nd ACM International Conference on Information and Knowledge Management}}. \bibinfo{pages}{4516--4522}.
\newblock


\bibitem[Cheng et~al\mbox{.}(2016)]%
        {cheng2016wd}
\bibfield{author}{\bibinfo{person}{Heng-Tze Cheng}, \bibinfo{person}{Levent Koc}, \bibinfo{person}{Jeremiah Harmsen}, \bibinfo{person}{Tal Shaked}, \bibinfo{person}{Tushar Chandra}, \bibinfo{person}{Hrishi Aradhye}, \bibinfo{person}{Glen Anderson}, \bibinfo{person}{Greg Corrado}, \bibinfo{person}{Wei Chai}, \bibinfo{person}{Mustafa Ispir}, {et~al\mbox{.}}} \bibinfo{year}{2016}\natexlab{}.
\newblock \showarticletitle{Wide \& deep learning for recommender systems}. In \bibinfo{booktitle}{\emph{Proceedings of the 1st workshop on deep learning for recommender systems}}. \bibinfo{pages}{7--10}.
\newblock


\bibitem[Dai et~al\mbox{.}(2023)]%
        {dai2023contrastive}
\bibfield{author}{\bibinfo{person}{Shitong Dai}, \bibinfo{person}{Jiongnan Liu}, \bibinfo{person}{Zhicheng Dou}, \bibinfo{person}{Haonan Wang}, \bibinfo{person}{Lin Liu}, \bibinfo{person}{Bo Long}, {and} \bibinfo{person}{Ji-Rong Wen}.} \bibinfo{year}{2023}\natexlab{}.
\newblock \showarticletitle{Contrastive Learning for User Sequence Representation in Personalized Product Search}. In \bibinfo{booktitle}{\emph{Proceedings of the 29th ACM SIGKDD Conference on Knowledge Discovery and Data Mining}}. \bibinfo{pages}{380--389}.
\newblock


\bibitem[Ge et~al\mbox{.}(2020)]%
        {ge2020understanding}
\bibfield{author}{\bibinfo{person}{Yingqiang Ge}, \bibinfo{person}{Shuya Zhao}, \bibinfo{person}{Honglu Zhou}, \bibinfo{person}{Changhua Pei}, \bibinfo{person}{Fei Sun}, \bibinfo{person}{Wenwu Ou}, {and} \bibinfo{person}{Yongfeng Zhang}.} \bibinfo{year}{2020}\natexlab{}.
\newblock \showarticletitle{Understanding echo chambers in e-commerce recommender systems}. In \bibinfo{booktitle}{\emph{Proceedings of the 43rd international ACM SIGIR conference on research and development in information retrieval}}. \bibinfo{pages}{2261--2270}.
\newblock


\bibitem[Geiger et~al\mbox{.}(2021)]%
        {geiger2021causal}
\bibfield{author}{\bibinfo{person}{Atticus Geiger}, \bibinfo{person}{Hanson Lu}, \bibinfo{person}{Thomas Icard}, {and} \bibinfo{person}{Christopher Potts}.} \bibinfo{year}{2021}\natexlab{}.
\newblock \showarticletitle{Causal abstractions of neural networks}.
\newblock \bibinfo{journal}{\emph{Advances in Neural Information Processing Systems}}  \bibinfo{volume}{34} (\bibinfo{year}{2021}), \bibinfo{pages}{9574--9586}.
\newblock


\bibitem[Geiger et~al\mbox{.}(2024)]%
        {geiger2024finding}
\bibfield{author}{\bibinfo{person}{Atticus Geiger}, \bibinfo{person}{Zhengxuan Wu}, \bibinfo{person}{Christopher Potts}, \bibinfo{person}{Thomas Icard}, {and} \bibinfo{person}{Noah Goodman}.} \bibinfo{year}{2024}\natexlab{}.
\newblock \showarticletitle{Finding alignments between interpretable causal variables and distributed neural representations}. In \bibinfo{booktitle}{\emph{Causal Learning and Reasoning}}. PMLR, \bibinfo{pages}{160--187}.
\newblock


\bibitem[Guo et~al\mbox{.}(2017)]%
        {guo2017deepfm}
\bibfield{author}{\bibinfo{person}{Huifeng Guo}, \bibinfo{person}{Ruiming Tang}, \bibinfo{person}{Yunming Ye}, \bibinfo{person}{Zhenguo Li}, {and} \bibinfo{person}{Xiuqiang He}.} \bibinfo{year}{2017}\natexlab{}.
\newblock \showarticletitle{DeepFM: a factorization-machine based neural network for CTR prediction}.
\newblock \bibinfo{journal}{\emph{arXiv preprint arXiv:1703.04247}} (\bibinfo{year}{2017}).
\newblock


\bibitem[Guo et~al\mbox{.}(2020)]%
        {guo2020survey}
\bibfield{author}{\bibinfo{person}{Ruocheng Guo}, \bibinfo{person}{Lu Cheng}, \bibinfo{person}{Jundong Li}, \bibinfo{person}{P~Richard Hahn}, {and} \bibinfo{person}{Huan Liu}.} \bibinfo{year}{2020}\natexlab{}.
\newblock \showarticletitle{A survey of learning causality with data: Problems and methods}.
\newblock \bibinfo{journal}{\emph{ACM Computing Surveys (CSUR)}} \bibinfo{volume}{53}, \bibinfo{number}{4} (\bibinfo{year}{2020}), \bibinfo{pages}{1--37}.
\newblock


\bibitem[Hong et~al\mbox{.}(2024)]%
        {hong2024print}
\bibfield{author}{\bibinfo{person}{Zhaolin Hong}, \bibinfo{person}{Haitao Wang}, \bibinfo{person}{Chengjie Qian}, \bibinfo{person}{Wei Chen}, \bibinfo{person}{Tianqi He}, \bibinfo{person}{Yajie Zou}, \bibinfo{person}{Qiang Liu}, {and} \bibinfo{person}{Xingxing Wang}.} \bibinfo{year}{2024}\natexlab{}.
\newblock \showarticletitle{PRINT: Personalized Relevance Incentive Network for CTR Prediction in Sponsored Search}. In \bibinfo{booktitle}{\emph{Companion Proceedings of the ACM on Web Conference 2024}}. \bibinfo{pages}{190--195}.
\newblock


\bibitem[Huang et~al\mbox{.}(2024)]%
        {huang2024utilizing}
\bibfield{author}{\bibinfo{person}{Jiahui Huang}, \bibinfo{person}{Lan Zhang}, \bibinfo{person}{Junhao Wang}, \bibinfo{person}{Shanyang Jiang}, \bibinfo{person}{Dongbo Huang}, \bibinfo{person}{Cheng Ding}, {and} \bibinfo{person}{Lan Xu}.} \bibinfo{year}{2024}\natexlab{}.
\newblock \showarticletitle{Utilizing Non-click Samples via Semi-supervised Learning for Conversion Rate Prediction}. In \bibinfo{booktitle}{\emph{Proceedings of the 18th ACM Conference on Recommender Systems}}. \bibinfo{pages}{350--359}.
\newblock


\bibitem[Huang et~al\mbox{.}(2013)]%
        {huang2013learning}
\bibfield{author}{\bibinfo{person}{Po-Sen Huang}, \bibinfo{person}{Xiaodong He}, \bibinfo{person}{Jianfeng Gao}, \bibinfo{person}{Li Deng}, \bibinfo{person}{Alex Acero}, {and} \bibinfo{person}{Larry Heck}.} \bibinfo{year}{2013}\natexlab{}.
\newblock \showarticletitle{Learning deep structured semantic models for web search using clickthrough data}. In \bibinfo{booktitle}{\emph{Proceedings of the 22nd ACM international conference on Information \& Knowledge Management}}. \bibinfo{pages}{2333--2338}.
\newblock


\bibitem[Jain et~al\mbox{.}(2021)]%
        {jain2021overview}
\bibfield{author}{\bibinfo{person}{Vipin Jain}, \bibinfo{person}{BINDOO Malviya}, {and} \bibinfo{person}{SATYENDRA Arya}.} \bibinfo{year}{2021}\natexlab{}.
\newblock \showarticletitle{An overview of electronic commerce (e-Commerce)}.
\newblock \bibinfo{journal}{\emph{The journal of contemporary issues in business and government}} \bibinfo{volume}{27}, \bibinfo{number}{3} (\bibinfo{year}{2021}), \bibinfo{pages}{665--670}.
\newblock


\bibitem[Jiang et~al\mbox{.}(2020)]%
        {jiang2020bert2dnn}
\bibfield{author}{\bibinfo{person}{Yunjiang Jiang}, \bibinfo{person}{Yue Shang}, \bibinfo{person}{Ziyang Liu}, \bibinfo{person}{Hongwei Shen}, \bibinfo{person}{Yun Xiao}, \bibinfo{person}{Sulong Xu}, \bibinfo{person}{Wei Xiong}, \bibinfo{person}{Weipeng Yan}, {and} \bibinfo{person}{Di Jin}.} \bibinfo{year}{2020}\natexlab{}.
\newblock \showarticletitle{BERT2DNN: BERT distillation with massive unlabeled data for online e-commerce search}. In \bibinfo{booktitle}{\emph{2020 IEEE International Conference on Data Mining (ICDM)}}. IEEE, \bibinfo{pages}{212--221}.
\newblock


\bibitem[Lee et~al\mbox{.}(2022)]%
        {lee2022bilateral}
\bibfield{author}{\bibinfo{person}{Jae-woong Lee}, \bibinfo{person}{Seongmin Park}, \bibinfo{person}{Joonseok Lee}, {and} \bibinfo{person}{Jongwuk Lee}.} \bibinfo{year}{2022}\natexlab{}.
\newblock \showarticletitle{Bilateral self-unbiased learning from biased implicit feedback}. In \bibinfo{booktitle}{\emph{Proceedings of the 45th International ACM SIGIR Conference on Research and Development in Information Retrieval}}. \bibinfo{pages}{29--39}.
\newblock


\bibitem[Li et~al\mbox{.}(2023a)]%
        {li2023strec}
\bibfield{author}{\bibinfo{person}{Chengxi Li}, \bibinfo{person}{Yejing Wang}, \bibinfo{person}{Qidong Liu}, \bibinfo{person}{Xiangyu Zhao}, \bibinfo{person}{Wanyu Wang}, \bibinfo{person}{Yiqi Wang}, \bibinfo{person}{Lixin Zou}, \bibinfo{person}{Wenqi Fan}, {and} \bibinfo{person}{Qing Li}.} \bibinfo{year}{2023}\natexlab{a}.
\newblock \showarticletitle{STRec: Sparse Transformer for Sequential Recommendations}. In \bibinfo{booktitle}{\emph{Proceedings of the 17th ACM Conference on Recommender Systems}}. \bibinfo{pages}{101--111}.
\newblock


\bibitem[Li et~al\mbox{.}(2022)]%
        {li2022gromov}
\bibfield{author}{\bibinfo{person}{Xinhang Li}, \bibinfo{person}{Zhaopeng Qiu}, \bibinfo{person}{Xiangyu Zhao}, \bibinfo{person}{Zihao Wang}, \bibinfo{person}{Yong Zhang}, \bibinfo{person}{Chunxiao Xing}, {and} \bibinfo{person}{Xian Wu}.} \bibinfo{year}{2022}\natexlab{}.
\newblock \showarticletitle{Gromov-wasserstein guided representation learning for cross-domain recommendation}. In \bibinfo{booktitle}{\emph{Proceedings of the 31st ACM International Conference on Information \& Knowledge Management}}. \bibinfo{pages}{1199--1208}.
\newblock


\bibitem[Li et~al\mbox{.}(2023b)]%
        {li2023hamur}
\bibfield{author}{\bibinfo{person}{Xiaopeng Li}, \bibinfo{person}{Fan Yan}, \bibinfo{person}{Xiangyu Zhao}, \bibinfo{person}{Yichao Wang}, \bibinfo{person}{Bo Chen}, \bibinfo{person}{Huifeng Guo}, {and} \bibinfo{person}{Ruiming Tang}.} \bibinfo{year}{2023}\natexlab{b}.
\newblock \showarticletitle{Hamur: Hyper adapter for multi-domain recommendation}. In \bibinfo{booktitle}{\emph{Proceedings of the 32nd ACM International Conference on Information and Knowledge Management}}. \bibinfo{pages}{1268--1277}.
\newblock


\bibitem[Li et~al\mbox{.}(2023c)]%
        {li2023imf}
\bibfield{author}{\bibinfo{person}{Xinhang Li}, \bibinfo{person}{Xiangyu Zhao}, \bibinfo{person}{Jiaxing Xu}, \bibinfo{person}{Yong Zhang}, {and} \bibinfo{person}{Chunxiao Xing}.} \bibinfo{year}{2023}\natexlab{c}.
\newblock \showarticletitle{IMF: Interactive Multimodal Fusion Model for Link Prediction}. In \bibinfo{booktitle}{\emph{Proceedings of the Web Conference 2023}}.
\newblock


\bibitem[Liang et~al\mbox{.}(2023)]%
        {liang2023mmmlp}
\bibfield{author}{\bibinfo{person}{Jiahao Liang}, \bibinfo{person}{Xiangyu Zhao}, \bibinfo{person}{Muyang Li}, \bibinfo{person}{Zijian Zhang}, \bibinfo{person}{Wanyu Wang}, \bibinfo{person}{Haochen Liu}, {and} \bibinfo{person}{Zitao Liu}.} \bibinfo{year}{2023}\natexlab{}.
\newblock \showarticletitle{Mmmlp: Multi-modal multilayer perceptron for sequential recommendations}. In \bibinfo{booktitle}{\emph{Proceedings of the ACM Web Conference 2023}}. \bibinfo{pages}{1109--1117}.
\newblock


\bibitem[Lin et~al\mbox{.}(2024)]%
        {lin2024recrec}
\bibfield{author}{\bibinfo{person}{Siyi Lin}, \bibinfo{person}{Sheng Zhou}, \bibinfo{person}{Jiawei Chen}, \bibinfo{person}{Yan Feng}, \bibinfo{person}{Qihao Shi}, \bibinfo{person}{Chun Chen}, \bibinfo{person}{Ying Li}, {and} \bibinfo{person}{Can Wang}.} \bibinfo{year}{2024}\natexlab{}.
\newblock \showarticletitle{ReCRec: Reasoning the Causes of Implicit Feedback for Debiased Recommendation}.
\newblock \bibinfo{journal}{\emph{ACM Transactions on Information Systems}} (\bibinfo{year}{2024}).
\newblock


\bibitem[Lin et~al\mbox{.}(2022)]%
        {lin2022adafs}
\bibfield{author}{\bibinfo{person}{Weilin Lin}, \bibinfo{person}{Xiangyu Zhao}, \bibinfo{person}{Yejing Wang}, \bibinfo{person}{Tong Xu}, {and} \bibinfo{person}{Xian Wu}.} \bibinfo{year}{2022}\natexlab{}.
\newblock \showarticletitle{AdaFS: Adaptive Feature Selection in Deep Recommender System}. In \bibinfo{booktitle}{\emph{Proceedings of the 28th ACM SIGKDD Conference on Knowledge Discovery and Data Mining}}. \bibinfo{pages}{3309--3317}.
\newblock


\bibitem[Lin et~al\mbox{.}(2023)]%
        {lin2023autodenoise}
\bibfield{author}{\bibinfo{person}{Weilin Lin}, \bibinfo{person}{Xiangyu Zhao}, \bibinfo{person}{Yejing Wang}, \bibinfo{person}{Yuanshao Zhu}, {and} \bibinfo{person}{Wanyu Wang}.} \bibinfo{year}{2023}\natexlab{}.
\newblock \showarticletitle{AutoDenoise: Automatic Data Instance Denoising for Recommendations}. In \bibinfo{booktitle}{\emph{Proceedings of the ACM Web Conference 2023}}. \bibinfo{pages}{1003--1011}.
\newblock


\bibitem[Liu et~al\mbox{.}(2022)]%
        {liu2022rating}
\bibfield{author}{\bibinfo{person}{Haochen Liu}, \bibinfo{person}{Da Tang}, \bibinfo{person}{Ji Yang}, \bibinfo{person}{Xiangyu Zhao}, \bibinfo{person}{Hui Liu}, \bibinfo{person}{Jiliang Tang}, {and} \bibinfo{person}{Youlong Cheng}.} \bibinfo{year}{2022}\natexlab{}.
\newblock \showarticletitle{Rating Distribution Calibration for Selection Bias Mitigation in Recommendations}. In \bibinfo{booktitle}{\emph{Proceedings of the ACM Web Conference 2022}}. \bibinfo{pages}{2048--2057}.
\newblock


\bibitem[Liu et~al\mbox{.}(2023b)]%
        {liu2023jdsearch}
\bibfield{author}{\bibinfo{person}{Jiongnan Liu}, \bibinfo{person}{Zhicheng Dou}, \bibinfo{person}{Guoyu Tang}, {and} \bibinfo{person}{Sulong Xu}.} \bibinfo{year}{2023}\natexlab{b}.
\newblock \showarticletitle{Jdsearch: A personalized product search dataset with real queries and full interactions}. In \bibinfo{booktitle}{\emph{Proceedings of the 46th International ACM SIGIR Conference on Research and Development in Information Retrieval}}. \bibinfo{pages}{2945--2952}.
\newblock


\bibitem[Liu et~al\mbox{.}(2023c)]%
        {liu2023multimodal}
\bibfield{author}{\bibinfo{person}{Qidong Liu}, \bibinfo{person}{Jiaxi Hu}, \bibinfo{person}{Yutian Xiao}, \bibinfo{person}{Xiangyu Zhao}, \bibinfo{person}{Jingtong Gao}, \bibinfo{person}{Wanyu Wang}, \bibinfo{person}{Qing Li}, {and} \bibinfo{person}{Jiliang Tang}.} \bibinfo{year}{2023}\natexlab{c}.
\newblock \showarticletitle{Multimodal recommender systems: A survey}.
\newblock \bibinfo{journal}{\emph{Comput. Surveys}} (\bibinfo{year}{2023}).
\newblock


\bibitem[Liu et~al\mbox{.}(2024a)]%
        {liu2024large}
\bibfield{author}{\bibinfo{person}{Qidong Liu}, \bibinfo{person}{Xian Wu}, \bibinfo{person}{Xiangyu Zhao}, \bibinfo{person}{Yejing Wang}, \bibinfo{person}{Zijian Zhang}, \bibinfo{person}{Feng Tian}, {and} \bibinfo{person}{Yefeng Zheng}.} \bibinfo{year}{2024}\natexlab{a}.
\newblock \showarticletitle{Large Language Models Enhanced Sequential Recommendation for Long-tail User and Item}.
\newblock \bibinfo{journal}{\emph{arXiv preprint arXiv:2405.20646}} (\bibinfo{year}{2024}).
\newblock


\bibitem[Liu et~al\mbox{.}(2024b)]%
        {liu2024moe}
\bibfield{author}{\bibinfo{person}{Qidong Liu}, \bibinfo{person}{Xian Wu}, \bibinfo{person}{Xiangyu Zhao}, \bibinfo{person}{Yuanshao Zhu}, \bibinfo{person}{Derong Xu}, \bibinfo{person}{Feng Tian}, {and} \bibinfo{person}{Yefeng Zheng}.} \bibinfo{year}{2024}\natexlab{b}.
\newblock \showarticletitle{When MOE Meets LLMs: Parameter Efficient Fine-tuning for Multi-task Medical Applications}. In \bibinfo{booktitle}{\emph{Proceedings of the 47th International ACM SIGIR Conference on Research and Development in Information Retrieval}}. \bibinfo{pages}{1104--1114}.
\newblock


\bibitem[Liu et~al\mbox{.}(2024c)]%
        {liu2024largesurvey}
\bibfield{author}{\bibinfo{person}{Qidong Liu}, \bibinfo{person}{Xiangyu Zhao}, \bibinfo{person}{Yuhao Wang}, \bibinfo{person}{Yejing Wang}, \bibinfo{person}{Zijian Zhang}, \bibinfo{person}{Yuqi Sun}, \bibinfo{person}{Xiang Li}, \bibinfo{person}{Maolin Wang}, \bibinfo{person}{Pengyue Jia}, \bibinfo{person}{Chong Chen}, {et~al\mbox{.}}} \bibinfo{year}{2024}\natexlab{c}.
\newblock \showarticletitle{Large Language Model Enhanced Recommender Systems: Taxonomy, Trend, Application and Future}.
\newblock \bibinfo{journal}{\emph{arXiv preprint arXiv:2412.13432}} (\bibinfo{year}{2024}).
\newblock


\bibitem[Liu et~al\mbox{.}(2023a)]%
        {liu2023exploration}
\bibfield{author}{\bibinfo{person}{Shuchang Liu}, \bibinfo{person}{Qingpeng Cai}, \bibinfo{person}{Bowen Sun}, \bibinfo{person}{Yuhao Wang}, \bibinfo{person}{Ji Jiang}, \bibinfo{person}{Dong Zheng}, \bibinfo{person}{Kun Gai}, \bibinfo{person}{Peng Jiang}, \bibinfo{person}{Xiangyu Zhao}, {and} \bibinfo{person}{Yongfeng Zhang}.} \bibinfo{year}{2023}\natexlab{a}.
\newblock \showarticletitle{Exploration and Regularization of the Latent Action Space in Recommendation}. In \bibinfo{booktitle}{\emph{Proceedings of the Web Conference 2023}}.
\newblock


\bibitem[Liu et~al\mbox{.}(2023d)]%
        {liu2023multi}
\bibfield{author}{\bibinfo{person}{Ziru Liu}, \bibinfo{person}{Jiejie Tian}, \bibinfo{person}{Qingpeng Cai}, \bibinfo{person}{Xiangyu Zhao}, \bibinfo{person}{Jingtong Gao}, \bibinfo{person}{Shuchang Liu}, \bibinfo{person}{Dayou Chen}, \bibinfo{person}{Tonghao He}, \bibinfo{person}{Dong Zheng}, \bibinfo{person}{Peng Jiang}, {et~al\mbox{.}}} \bibinfo{year}{2023}\natexlab{d}.
\newblock \showarticletitle{Multi-task recommendations with reinforcement learning}. In \bibinfo{booktitle}{\emph{Proceedings of the ACM Web Conference 2023}}. \bibinfo{pages}{1273--1282}.
\newblock


\bibitem[Nguyen et~al\mbox{.}(2020)]%
        {nguyen2020learning}
\bibfield{author}{\bibinfo{person}{Thanh Nguyen}, \bibinfo{person}{Nikhil Rao}, {and} \bibinfo{person}{Karthik Subbian}.} \bibinfo{year}{2020}\natexlab{}.
\newblock \showarticletitle{Learning Robust Models for e-Commerce Product Search}. In \bibinfo{booktitle}{\emph{Proceedings of the 58th Annual Meeting of the Association for Computational Linguistics}}. \bibinfo{pages}{6861--6869}.
\newblock


\bibitem[Robertson et~al\mbox{.}(2009)]%
        {robertson2009probabilistic}
\bibfield{author}{\bibinfo{person}{Stephen Robertson}, \bibinfo{person}{Hugo Zaragoza}, {et~al\mbox{.}}} \bibinfo{year}{2009}\natexlab{}.
\newblock \showarticletitle{The probabilistic relevance framework: BM25 and beyond}.
\newblock \bibinfo{journal}{\emph{Foundations and Trends{\textregistered} in Information Retrieval}} \bibinfo{volume}{3}, \bibinfo{number}{4} (\bibinfo{year}{2009}), \bibinfo{pages}{333--389}.
\newblock


\bibitem[Shi et~al\mbox{.}(2024)]%
        {shi2024unisar}
\bibfield{author}{\bibinfo{person}{Teng Shi}, \bibinfo{person}{Zihua Si}, \bibinfo{person}{Jun Xu}, \bibinfo{person}{Xiao Zhang}, \bibinfo{person}{Xiaoxue Zang}, \bibinfo{person}{Kai Zheng}, \bibinfo{person}{Dewei Leng}, \bibinfo{person}{Yanan Niu}, {and} \bibinfo{person}{Yang Song}.} \bibinfo{year}{2024}\natexlab{}.
\newblock \showarticletitle{UniSAR: Modeling User Transition Behaviors between Search and Recommendation}. In \bibinfo{booktitle}{\emph{Proceedings of the 47th International ACM SIGIR Conference on Research and Development in Information Retrieval}}. \bibinfo{pages}{1029--1039}.
\newblock


\bibitem[Sun et~al\mbox{.}(2023)]%
        {sun2023kuaisar}
\bibfield{author}{\bibinfo{person}{Zhongxiang Sun}, \bibinfo{person}{Zihua Si}, \bibinfo{person}{Xiaoxue Zang}, \bibinfo{person}{Dewei Leng}, \bibinfo{person}{Yanan Niu}, \bibinfo{person}{Yang Song}, \bibinfo{person}{Xiao Zhang}, {and} \bibinfo{person}{Jun Xu}.} \bibinfo{year}{2023}\natexlab{}.
\newblock \showarticletitle{KuaiSAR: A Unified Search And Recommendation Dataset}. In \bibinfo{booktitle}{\emph{Proceedings of the 32nd ACM International Conference on Information and Knowledge Management}}. \bibinfo{pages}{5407--5411}.
\newblock


\bibitem[Wang et~al\mbox{.}(2022a)]%
        {wang2022escm2}
\bibfield{author}{\bibinfo{person}{Hao Wang}, \bibinfo{person}{Tai-Wei Chang}, \bibinfo{person}{Tianqiao Liu}, \bibinfo{person}{Jianmin Huang}, \bibinfo{person}{Zhichao Chen}, \bibinfo{person}{Chao Yu}, \bibinfo{person}{Ruopeng Li}, {and} \bibinfo{person}{Wei Chu}.} \bibinfo{year}{2022}\natexlab{a}.
\newblock \showarticletitle{ESCM2: entire space counterfactual multi-task model for post-click conversion rate estimation}. In \bibinfo{booktitle}{\emph{Proceedings of the 45th International ACM SIGIR Conference on Research and Development in Information Retrieval}}. \bibinfo{pages}{363--372}.
\newblock


\bibitem[Wang et~al\mbox{.}(2017)]%
        {wang2017dcn}
\bibfield{author}{\bibinfo{person}{Ruoxi Wang}, \bibinfo{person}{Bin Fu}, \bibinfo{person}{Gang Fu}, {and} \bibinfo{person}{Mingliang Wang}.} \bibinfo{year}{2017}\natexlab{}.
\newblock \showarticletitle{Deep \& cross network for ad click predictions}.
\newblock In \bibinfo{booktitle}{\emph{Proceedings of the ADKDD'17}}. \bibinfo{pages}{1--7}.
\newblock


\bibitem[Wang et~al\mbox{.}(2023a)]%
        {wang2023single}
\bibfield{author}{\bibinfo{person}{Yejing Wang}, \bibinfo{person}{Zhaocheng Du}, \bibinfo{person}{Xiangyu Zhao}, \bibinfo{person}{Bo Chen}, \bibinfo{person}{Huifeng Guo}, \bibinfo{person}{Ruiming Tang}, {and} \bibinfo{person}{Zhenhua Dong}.} \bibinfo{year}{2023}\natexlab{a}.
\newblock \showarticletitle{Single-shot Feature Selection for Multi-task Recommendations}. In \bibinfo{booktitle}{\emph{Proceedings of the 46th International ACM SIGIR Conference on Research and Development in Information Retrieval}}. \bibinfo{pages}{341--351}.
\newblock


\bibitem[Wang et~al\mbox{.}(2023b)]%
        {wang2023doctor}
\bibfield{author}{\bibinfo{person}{Yejing Wang}, \bibinfo{person}{Shen Ge}, \bibinfo{person}{Xiangyu Zhao}, \bibinfo{person}{Xian Wu}, \bibinfo{person}{Tong Xu}, \bibinfo{person}{Chen Ma}, {and} \bibinfo{person}{Zhi Zheng}.} \bibinfo{year}{2023}\natexlab{b}.
\newblock \showarticletitle{Doctor specific tag recommendation for online medical record management}. In \bibinfo{booktitle}{\emph{Proceedings of the 29th ACM SIGKDD Conference on Knowledge Discovery and Data Mining}}. \bibinfo{pages}{5150--5161}.
\newblock


\bibitem[Wang et~al\mbox{.}(2024)]%
        {wang2024gprec}
\bibfield{author}{\bibinfo{person}{Yejing Wang}, \bibinfo{person}{Dong Xu}, \bibinfo{person}{Xiangyu Zhao}, \bibinfo{person}{Zhiren Mao}, \bibinfo{person}{Peng Xiang}, \bibinfo{person}{Ling Yan}, \bibinfo{person}{Yao Hu}, \bibinfo{person}{Zijian Zhang}, \bibinfo{person}{Xuetao Wei}, {and} \bibinfo{person}{Qidong Liu}.} \bibinfo{year}{2024}\natexlab{}.
\newblock \showarticletitle{GPRec: Bi-level User Modeling for Deep Recommenders}.
\newblock \bibinfo{journal}{\emph{arXiv preprint arXiv:2410.20730}} (\bibinfo{year}{2024}).
\newblock


\bibitem[Wang et~al\mbox{.}(2023c)]%
        {wang2023plate}
\bibfield{author}{\bibinfo{person}{Yuhao Wang}, \bibinfo{person}{Xiangyu Zhao}, \bibinfo{person}{Bo Chen}, \bibinfo{person}{Qidong Liu}, \bibinfo{person}{Huifeng Guo}, \bibinfo{person}{Huanshuo Liu}, \bibinfo{person}{Yichao Wang}, \bibinfo{person}{Rui Zhang}, {and} \bibinfo{person}{Ruiming Tang}.} \bibinfo{year}{2023}\natexlab{c}.
\newblock \showarticletitle{PLATE: A Prompt-Enhanced Paradigm for Multi-Scenario Recommendations}. In \bibinfo{booktitle}{\emph{Proceedings of the 46th International ACM SIGIR Conference on Research and Development in Information Retrieval}}. \bibinfo{pages}{1498--1507}.
\newblock


\bibitem[Wang et~al\mbox{.}(2022c)]%
        {wang2022autofield}
\bibfield{author}{\bibinfo{person}{Yejing Wang}, \bibinfo{person}{Xiangyu Zhao}, \bibinfo{person}{Tong Xu}, {and} \bibinfo{person}{Xian Wu}.} \bibinfo{year}{2022}\natexlab{c}.
\newblock \showarticletitle{Autofield: Automating feature selection in deep recommender systems}. In \bibinfo{booktitle}{\emph{Proceedings of the ACM Web Conference 2022}}. \bibinfo{pages}{1977--1986}.
\newblock


\bibitem[Wang et~al\mbox{.}(2022b)]%
        {wang2022invpref}
\bibfield{author}{\bibinfo{person}{Zimu Wang}, \bibinfo{person}{Yue He}, \bibinfo{person}{Jiashuo Liu}, \bibinfo{person}{Wenchao Zou}, \bibinfo{person}{Philip~S Yu}, {and} \bibinfo{person}{Peng Cui}.} \bibinfo{year}{2022}\natexlab{b}.
\newblock \showarticletitle{Invariant preference learning for general debiasing in recommendation}. In \bibinfo{booktitle}{\emph{Proceedings of the 28th ACM SIGKDD Conference on Knowledge Discovery and Data Mining}}. \bibinfo{pages}{1969--1978}.
\newblock


\bibitem[Wang et~al\mbox{.}(2021)]%
        {wang2021masknet}
\bibfield{author}{\bibinfo{person}{Zhiqiang Wang}, \bibinfo{person}{Qingyun She}, {and} \bibinfo{person}{Junlin Zhang}.} \bibinfo{year}{2021}\natexlab{}.
\newblock \showarticletitle{MaskNet: Introducing Feature-Wise Multiplication to CTR Ranking Models by Instance-Guided Mask}.
\newblock  (\bibinfo{year}{2021}).
\newblock


\bibitem[Wang~Yuxin(2023)]%
        {Moka}
\bibfield{author}{\bibinfo{person}{He~sicheng Wang~Yuxin, Sun~Qingxuan}.} \bibinfo{year}{2023}\natexlab{}.
\newblock \bibinfo{booktitle}{\emph{Moka Massive Mixed Embedding}}.
\newblock
\urldef\tempurl%
\url{https://huggingface.co/moka-ai/m3e-base}
\showURL{%
\tempurl}


\bibitem[Xiao et~al\mbox{.}(2019)]%
        {xiao2019weakly}
\bibfield{author}{\bibinfo{person}{Rong Xiao}, \bibinfo{person}{Jianhui Ji}, \bibinfo{person}{Baoliang Cui}, \bibinfo{person}{Haihong Tang}, \bibinfo{person}{Wenwu Ou}, \bibinfo{person}{Yanghua Xiao}, \bibinfo{person}{Jiwei Tan}, {and} \bibinfo{person}{Xuan Ju}.} \bibinfo{year}{2019}\natexlab{}.
\newblock \showarticletitle{Weakly supervised co-training of query rewriting and semantic matching for e-commerce}. In \bibinfo{booktitle}{\emph{Proceedings of the twelfth ACM international conference on web search and data mining}}. \bibinfo{pages}{402--410}.
\newblock


\bibitem[Xu et~al\mbox{.}(2024)]%
        {xu2024large}
\bibfield{author}{\bibinfo{person}{Derong Xu}, \bibinfo{person}{Wei Chen}, \bibinfo{person}{Wenjun Peng}, \bibinfo{person}{Chao Zhang}, \bibinfo{person}{Tong Xu}, \bibinfo{person}{Xiangyu Zhao}, \bibinfo{person}{Xian Wu}, \bibinfo{person}{Yefeng Zheng}, \bibinfo{person}{Yang Wang}, {and} \bibinfo{person}{Enhong Chen}.} \bibinfo{year}{2024}\natexlab{}.
\newblock \showarticletitle{Large language models for generative information extraction: A survey}.
\newblock \bibinfo{journal}{\emph{Frontiers of Computer Science}} \bibinfo{volume}{18}, \bibinfo{number}{6} (\bibinfo{year}{2024}), \bibinfo{pages}{186357}.
\newblock


\bibitem[Yao et~al\mbox{.}(2021)]%
        {yao2021learningpositionbias}
\bibfield{author}{\bibinfo{person}{Shaowei Yao}, \bibinfo{person}{Jiwei Tan}, \bibinfo{person}{Xi Chen}, \bibinfo{person}{Keping Yang}, \bibinfo{person}{Rong Xiao}, \bibinfo{person}{Hongbo Deng}, {and} \bibinfo{person}{Xiaojun Wan}.} \bibinfo{year}{2021}\natexlab{}.
\newblock \showarticletitle{Learning a product relevance model from click-through data in e-commerce}. In \bibinfo{booktitle}{\emph{Proceedings of the Web Conference 2021}}. \bibinfo{pages}{2890--2899}.
\newblock


\bibitem[Zan et~al\mbox{.}(2023)]%
        {zan2023spm}
\bibfield{author}{\bibinfo{person}{Wen Zan}, \bibinfo{person}{Yaopeng Han}, \bibinfo{person}{Xiaotian Jiang}, \bibinfo{person}{Yao Xiao}, \bibinfo{person}{Yang Yang}, \bibinfo{person}{Dayao Chen}, {and} \bibinfo{person}{Sheng Chen}.} \bibinfo{year}{2023}\natexlab{}.
\newblock \showarticletitle{SPM: Structured Pretraining and Matching Architectures for Relevance Modeling in Meituan Search}. In \bibinfo{booktitle}{\emph{Proceedings of the 32nd ACM International Conference on Information and Knowledge Management}}. \bibinfo{pages}{4923--4929}.
\newblock


\bibitem[Zhang et~al\mbox{.}(2022)]%
        {zhang2022hierarchical}
\bibfield{author}{\bibinfo{person}{Chi Zhang}, \bibinfo{person}{Yantong Du}, \bibinfo{person}{Xiangyu Zhao}, \bibinfo{person}{Qilong Han}, \bibinfo{person}{Rui Chen}, {and} \bibinfo{person}{Li Li}.} \bibinfo{year}{2022}\natexlab{}.
\newblock \showarticletitle{Hierarchical Item Inconsistency Signal Learning for Sequence Denoising in Sequential Recommendation}. In \bibinfo{booktitle}{\emph{Proceedings of the 31st ACM International Conference on Information \& Knowledge Management}}. \bibinfo{pages}{2508--2518}.
\newblock


\bibitem[Zhang et~al\mbox{.}(2016)]%
        {zhang2016ffn}
\bibfield{author}{\bibinfo{person}{Weinan Zhang}, \bibinfo{person}{Tianming Du}, {and} \bibinfo{person}{Jun Wang}.} \bibinfo{year}{2016}\natexlab{}.
\newblock \showarticletitle{Deep Learning over Multi-field Categorical Data: --A Case Study on User Response Prediction}. In \bibinfo{booktitle}{\emph{Advances in Information Retrieval: 38th European Conference on IR Research, ECIR 2016, Padua, Italy, March 20--23, 2016. Proceedings 38}}. Springer, \bibinfo{pages}{45--57}.
\newblock


\bibitem[Zhao et~al\mbox{.}(2023)]%
        {zhao2023embedding}
\bibfield{author}{\bibinfo{person}{Xiangyu Zhao}, \bibinfo{person}{Maolin Wang}, \bibinfo{person}{Xinjian Zhao}, \bibinfo{person}{Jiansheng Li}, \bibinfo{person}{Shucheng Zhou}, \bibinfo{person}{Dawei Yin}, \bibinfo{person}{Qing Li}, \bibinfo{person}{Jiliang Tang}, {and} \bibinfo{person}{Ruocheng Guo}.} \bibinfo{year}{2023}\natexlab{}.
\newblock \showarticletitle{Embedding in recommender systems: A survey}.
\newblock \bibinfo{journal}{\emph{arXiv preprint arXiv:2310.18608}} (\bibinfo{year}{2023}).
\newblock


\bibitem[Zhao et~al\mbox{.}(2018a)]%
        {zhao2018deep}
\bibfield{author}{\bibinfo{person}{Xiangyu Zhao}, \bibinfo{person}{Long Xia}, \bibinfo{person}{Liang Zhang}, \bibinfo{person}{Zhuoye Ding}, \bibinfo{person}{Dawei Yin}, {and} \bibinfo{person}{Jiliang Tang}.} \bibinfo{year}{2018}\natexlab{a}.
\newblock \showarticletitle{Deep Reinforcement Learning for Page-wise Recommendations}. In \bibinfo{booktitle}{\emph{Proceedings of the 12th ACM Recommender Systems Conference}}. ACM, \bibinfo{pages}{95--103}.
\newblock


\bibitem[Zhao et~al\mbox{.}(2020)]%
        {zhao2020whole}
\bibfield{author}{\bibinfo{person}{Xiangyu Zhao}, \bibinfo{person}{Long Xia}, \bibinfo{person}{Lixin Zou}, \bibinfo{person}{Hui Liu}, \bibinfo{person}{Dawei Yin}, {and} \bibinfo{person}{Jiliang Tang}.} \bibinfo{year}{2020}\natexlab{}.
\newblock \showarticletitle{Whole-Chain Recommendations}. In \bibinfo{booktitle}{\emph{Proceedings of the 29th ACM International Conference on Information \& Knowledge Management}}. \bibinfo{pages}{1883--1891}.
\newblock


\bibitem[Zhao et~al\mbox{.}(2018b)]%
        {zhao2018recommendations}
\bibfield{author}{\bibinfo{person}{Xiangyu Zhao}, \bibinfo{person}{Liang Zhang}, \bibinfo{person}{Zhuoye Ding}, \bibinfo{person}{Long Xia}, \bibinfo{person}{Jiliang Tang}, {and} \bibinfo{person}{Dawei Yin}.} \bibinfo{year}{2018}\natexlab{b}.
\newblock \showarticletitle{Recommendations with Negative Feedback via Pairwise Deep Reinforcement Learning}. In \bibinfo{booktitle}{\emph{Proceedings of the 24th ACM SIGKDD International Conference on Knowledge Discovery \& Data Mining}}. ACM, \bibinfo{pages}{1040--1048}.
\newblock


\bibitem[Zheng et~al\mbox{.}(2022)]%
        {zheng2022cbr}
\bibfield{author}{\bibinfo{person}{Zhi Zheng}, \bibinfo{person}{Zhaopeng Qiu}, \bibinfo{person}{Tong Xu}, \bibinfo{person}{Xian Wu}, \bibinfo{person}{Xiangyu Zhao}, \bibinfo{person}{Enhong Chen}, {and} \bibinfo{person}{Hui Xiong}.} \bibinfo{year}{2022}\natexlab{}.
\newblock \showarticletitle{CBR: Context Bias aware Recommendation for Debiasing User Modeling and Click Prediction}. In \bibinfo{booktitle}{\emph{Proceedings of the ACM Web Conference 2022}}. \bibinfo{pages}{2268--2276}.
\newblock


\bibitem[Zhu et~al\mbox{.}(2023)]%
        {zhu2023dcmt}
\bibfield{author}{\bibinfo{person}{Feng Zhu}, \bibinfo{person}{Mingjie Zhong}, \bibinfo{person}{Xinxing Yang}, \bibinfo{person}{Longfei Li}, \bibinfo{person}{Lu Yu}, \bibinfo{person}{Tiehua Zhang}, \bibinfo{person}{Jun Zhou}, \bibinfo{person}{Chaochao Chen}, \bibinfo{person}{Fei Wu}, \bibinfo{person}{Guanfeng Liu}, {et~al\mbox{.}}} \bibinfo{year}{2023}\natexlab{}.
\newblock \showarticletitle{DCMT: A Direct Entire-Space Causal Multi-Task Framework for Post-Click Conversion Estimation}. In \bibinfo{booktitle}{\emph{2023 IEEE 39th International Conference on Data Engineering (ICDE)}}. IEEE, \bibinfo{pages}{3113--3125}.
\newblock


\end{thebibliography}

\appendix

\section{Additional Results}\label{appsec:res}
\subsection{Comparison on DeepFM \& DESTINE}
\begin{table}[t]
\centering
\caption{Additional overall comparison on Private.}
\label{tab:app-ov}
\resizebox{\columnwidth}{!}{%
\begin{tabular}{ccccccc}
\toprule
\textbf{PM}                & \textbf{RM} & \textbf{JM} & \textbf{AUC}    & \textbf{Logloss} & \textbf{NDCG}   & \textbf{HR}     \\\midrule
\multirow{6}{*}{DeepFM}  & DSSM      & Base      & 0.6619          & 0.05606          & 0.1581          & 0.2820          \\
                         &           & DRP       & \textbf{0.6631} & \textbf{0.05578} & \textbf{0.1623} & \textbf{0.2867} \\
                         & QEM       & Base      & 0.6658          & 0.05575          & 0.1635          & 0.2885          \\
                         &           & DRP       & \textbf{0.6707} & \textbf{0.05558} & \textbf{0.1658} & \textbf{0.2919} \\
                         & HEM       & Base      & 0.6661          & 0.05574          & 0.1633          & 0.2878          \\
                         &           & DRP       & \textbf{0.6697} & \textbf{0.05562} & \textbf{0.1650} & \textbf{0.2901} \\\midrule
\multirow{6}{*}{DESTINE} & DSSM      & Base      & 0.6709          & \textbf{0.05565} & 0.1660          & 0.2921          \\
                         &           & DRP       & \textbf{0.6715} & 0.05568          & \textbf{0.1671} & \textbf{0.2936} \\
                         & QEM       & Base      & 0.6697          & 0.05563          & 0.1657          & 0.2910          \\
                         &           & DRP       & \textbf{0.6709} & \textbf{0.05554} & \textbf{0.1668} & \textbf{0.2924} \\
                         & HEM       & Base      & 0.6706          & 0.05564          & 0.1662          & 0.2919          \\
                         &           & DRP       & \textbf{0.6711} & \textbf{0.05557} & \textbf{0.1676} & \textbf{0.2938}
\\\bottomrule
\end{tabular}%
}
\end{table}
We present performance of \name with preference model DeepFM and DESTINE on Private in Table~\ref{tab:app-ov}. Our findings indicate that DRP outperforms `Base' when using these PM backbones.

\subsection{Comparison with Disentanglement Method}
\begin{table}[t]
\centering
\caption{Additional disentanglement baseline on KuaiSAR.}
\label{tab:app-dis}
\resizebox{\columnwidth}{!}{%
\begin{tabular}{ccccccc}
\toprule
\textbf{PM} & \textbf{RM} & \textbf{JM} & \textbf{AUC}    & \textbf{Logloss} & \textbf{NDCG}   & \textbf{HR}     \\\midrule
\multirow{6}{*}{MLP}       & DSSM      & InvPref   & 0.6211          & 0.3554           & 0.4114          & 0.7117          \\
          &           & DRP       & \textbf{0.6229} & \textbf{0.3549}  & \textbf{0.4168} & \textbf{0.7145} \\
          & QEM       & InvPref   & 0.6122          & 0.3621           & 0.4101          & \textbf{0.7148} \\
          &           & DRP       & \textbf{0.6227} & \textbf{0.3551}  & \textbf{0.4143} & 0.7132          \\
          & HEM       & InvPref   & 0.6218          & 0.3552           & 0.4110          & 0.7121          \\
          &           & DRP       & \textbf{0.6230} & \textbf{0.3551}  & \textbf{0.4138} & \textbf{0.7122} \\\midrule
\multirow{6}{*}{DCN}      & DSSM      & InvPref   & \textbf{0.6206} & \textbf{0.3556}  & 0.4131          & 0.7112          \\
          &           & DRP       & 0.6200          & 0.3557           & \textbf{0.4145} & \textbf{0.7133} \\
          & QEM       & InvPref   & 0.6200          & 0.3555           & 0.4125          & 0.7134          \\
          &           & DRP       & \textbf{0.6243} & \textbf{0.3547}  & \textbf{0.4163} & \textbf{0.7142} \\
          & HEM       & InvPref   & 0.6223          & 0.3550           & 0.4136          & 0.7134          \\
          &           & DRP       & \textbf{0.6236} & \textbf{0.3545}  & \textbf{0.4152} & \textbf{0.7144}\\\bottomrule
\end{tabular}%
}
\end{table}
We compare the proposed \name with the latest disentanglement method InvPref~\cite{wang2022invpref} on KuaiSAR. The results is presented in Table~\ref{tab:app-dis}. DRP outperforms InvPref in most scenarios. However, InvPref demonstrates stronger compatibility when compared to other baselines in Table~\ref{tab:OV}. 

\end{sloppypar}
\end{document}